\documentclass[reqno,a4paper,final]{amsart}
\usepackage{amssymb,amstext,amsthm,eucal,amscd}
\usepackage{bbold}
\usepackage{array}
\usepackage[latin1]{inputenc}
\usepackage[notref,notcite]{showkeys}
\usepackage[vcentermath]{youngtab}
\Yboxdim{6pt}
\usepackage{hyperref}
\hypersetup{%
  pdftitle   = {Homogeneity and plane-wave limits},
  pdfkeywords = {supergravity, Killing spinors, isometries,
    homogeneity, supersymmetry, Penrose limits, plane-wave limits},
  pdfauthor  = {José Figueroa-O'Farrill, Patrick Meessen, Simon Philip},
  pdfcreator = {\LaTeX\ with package \flqq hyperref\frqq}
}
\newcommand{\half}{\tfrac{1}{2}}
\renewcommand{\d}{\partial}
\newcommand{\fa}{\mathfrak{a}}

\newcommand{\fg}{\mathfrak{g}}
\newcommand{\fh}{\mathfrak{h}}
\newcommand{\fk}{\mathfrak{k}}
\newcommand{\fm}{\mathfrak{m}}

\newcommand{\fz}{\mathfrak{z}}

\newcommand{\fiso}{\mathfrak{iso}}

\newcommand{\fso}{\mathfrak{so}}

\newcommand{\fsu}{\mathfrak{su}}
\newcommand{\fu}{\mathfrak{u}}
\renewcommand{\H}{\mathrm{H}}
\newcommand{\ISO}{\mathrm{ISO}}
\newcommand{\SO}{\mathrm{SO}}

\ifx\Sp\UnDeFiNeD
\newcommand{\Sp}{\mathrm{Sp}}
\else
\renewcommand{\Sp}{\mathrm{Sp}}
\fi

\newcommand{\SU}{\mathrm{SU}}
\newcommand{\U}{\mathrm{U}}
\newcommand{\EE}{\mathbb{E}}
\newcommand{\RR}{\mathbb{R}}
\newcommand{\CC}{\mathbb{C}}

\newcommand{\eT}{\mathcal{T}}

\DeclareMathOperator{\Ad}{Ad}

\DeclareMathOperator{\AdS}{AdS}

\DeclareMathOperator{\End}{End}

\newcommand{\bbee}{\boldsymbol{b}}
\newcommand{\be}{\boldsymbol{e}}
\newcommand{\bv}{\boldsymbol{v}}
\newcommand{\bw}{\boldsymbol{w}}
\newcommand{\bx}{\boldsymbol{x}}

\newcommand{\bzero}{\boldsymbol{0}}
\newcommand{\btheta}{\boldsymbol{\theta}}

\newcommand{\1}{\mathbb{1}}

\newcommand{\MUNCH}[1]{\relax}

\allowdisplaybreaks[1]
\begin{document}
\dedicatory{In memory of Stanley E. Hobert}
\title{Homogeneity and plane-wave limits}
\author[Figueroa-O'Farrill]{José Figueroa-O'Farrill}
\author[Meessen]{Patrick Meessen}
\author[Philip]{Simon Philip}
\address[JMF,SP]{School of Mathematics, University of Edinburgh,
  Scotland, UK}
\email{J.M.Figueroa@ed.ac.uk, S.A.R.Philip@sms.ed.ac.uk}
\address[PM]{Physics Department, Theory, CERN, Geneva, Switzerland}
\email{Patrick.Meessen@cern.ch}
\date{\today}
\thanks{EMPG-05-03; CERN-PH-TH/2005-057}
\begin{abstract}
  We explore the plane-wave limit of homogeneous spacetimes.  For
  plane-wave limits along homogeneous geodesics the limit is known to
  be homogeneous and we exhibit the limiting metric in terms of Lie
  algebraic data.  This simplifies many calculations and we illustrate
  this with several examples.  We also investigate the behaviour of
  (reductive) homogeneous structures under the plane-wave limit.
\end{abstract}
\maketitle
\tableofcontents

\section{Introduction and conclusion}
\label{sec:intro}

We have recently shown in \cite{FMPHom} that M-theory backgrounds
preserving more than 24 of the 32 supersymmetries are necessarily
(locally) homogeneous.  To date the only 24+ solutions known in the
literature are actually symmetric: Minkowski spacetime, the
Freund--Rubin backgrounds $\AdS_4\times S^7$ and $\AdS_7 \times S^4$,
the Kowalski-Glikman wave \cite{KG,FOPflux} and an Hpp-wave found by
Michelson \cite{Michelson26}.  Since \emph{all} maximally
supersymmetric backgrounds are (locally) symmetric, it is not
inconceivable that this might be forced by having less than maximal
(say, 24?+) supersymmetry.  If so, then this would allow a
classification of 24?+ solutions by suitably extending results already
in the literature.  In the absence of such a result, however, we are
forced to find other, more accessible characterisations of these
backgrounds.

One property of 24+ backgrounds is that all their plane-wave limits
are homogeneous.  This can be established by observing that since the
plane-wave limit can only enhance the preserved supersymmetry
\cite{Limits}, any plane-wave limit of a 24+ background must itself be
24+ and hence (locally) homogeneous by \cite{FMPHom}.  This is not a
trivial statement, because homogeneity is not always preserved under
the plane-wave limit, as illustrated by Patricot \cite{Patricot}, who
observed that the product of a Kaigorodov spacetime with a sphere
(which are homogeneous) admits nonhomogeneous plane-wave limits.  (In
fact, already the Kaigorodov space admits nonhomogeneous plane-wave
limits.)  Notice however that the plane-wave limit of a (locally)
symmetric space is (locally) symmetric; hence homogeneity is preserved
in these cases as well.  A natural question is then: What are
sufficient and necessary conditions that guarantee the preservation of
homogeneity under a plane-wave limit?

Given that the data for the plane-wave limit consists of both a
spacetime \emph{and} a null geodesic on it, it should not be
surprising that the homogeneity of the plane-wave limit depends on the
geodesic along which the limit is taken and not just on the spacetime.
Indeed, in \cite{SimonPL} it is shown that the limit is homogeneous if
it is taken along a homogeneous geodesic; that is, along the orbit of
a one-parameter subgroup of isometries.  This result has a clear
interpretation.  First of all, it is known that the isometry Lie
algebra undergoes a contraction in the limit, but in general we cannot
say whether the resulting algebra acts transitively or not on the
resulting plane wave.  The condition that the null geodesic be
homogeneous then means that we force some Killing vector to remain in
the translational part of the algebra.  Since plane waves are
generally of cohomogeneity one, this extra Killing vector guarantees
homogeneity.

In view of this, it is very tempting to conjecture that the plane-wave
limit of a homogeneous space is homogeneous if and only if the null
geodesic is homogeneous.  Even if this is not the case, the
homogeneous spaces on which all null geodesics are homogeneous, are
natural candidates for 24+ solutions and do merit further
investigation.

In this paper we will explore the interaction between homogeneity and
plane-wave limits.  Section~\ref{sec:homogeneity} is preparatory and
reviews the basic technology of homogeneous manifolds, homogeneous
structures, homogeneous geodesics as well as the classification of
homogeneous plane waves by Blau and O'Loughlin \cite{BOLhpw}.  Such
plane waves come in two classes and are characterised by certain
algebraic data, namely a real number and a symmetric and a
skew-symmetric bilinear forms.  In Section~\ref{sec:homogPL} we show
how to compute these from the Lie algebraic data associated to a
homogeneous spacetime and a homogeneous geodesic, in effect reducing
the computation of the plane-wave limit to an algebraic calculation
which can be easily implemented in the computer, for example.  We
present two derivations of this result: one using the covariant
description \cite{CovariantPL} of the plane-wave limit and one
involving a different limiting procedure equivalent, as we will show,
to the plane-wave limit but without the need to find neither adapted
frames nor adapted coordinates.  We will also study the behaviour of
the homogeneous structures under the plane-wave limit.  Of course, the
plane-wave limit depends on the geodesic and not just on the
homogeneous structure, whence the results in this section are of
necessity somewhat less comprehensive.  Finally, in
Section~\ref{sec:examples} we present some examples to illustrate our
methods.

\section{Homogeneity}
\label{sec:homogeneity}

In this section we review the basic paraphernalia of homogeneous
manifolds and reductive homogeneous structures.

\subsection{Basic notions}
\label{sec:basic}

A pseudo-riemannian manifold $(M,g)$ is \textbf{homogeneous} if some
Lie group $G$ acts transitively on $M$ preserving the metric.  Closely
related to the notion of homogeneity is the notion of local
homogeneity.  This notion is important in the context of
(super)gravitational backgrounds, since often we are only dealing with
local metrics.  A manifold $(M,g)$ is \textbf{locally homogeneous} if
given any two points $p,q\in M$ there are neighbourhoods $U \ni p$ and
$V \ni q$ and a local isometry $f: U \to V$ such that $f(p) = q$.  The
crucial difference is that the isometry $f$ need not extend to all of
$M$.  For example, the sphere is homogeneous, but the sphere without
the North pole, say, is only locally homogeneous.  The isometries
which are defined on the sphere without the pole are those isometries
which fix the pole, and these only have the parallels as orbits.

For simplicity of exposition, let us consider the case of a
homogeneous manifold, so that we do have a transitive action of some
group of isometries.  Fixing a point $o\in M$, the smooth map
\begin{equation*}
  \phi_o: G \to M~,
\end{equation*}
sending a group element $x\in G$ to its action $x \cdot o$ on the
point, is surjective.  The subgroup $H \subset G$ which fixes the
point $o$ is called the \textbf{isotropy subgroup} of $o$.  The map
$\phi_o$ induces a diffeomorphism $G/H \cong M$, which allows us to
identify $M$ with the space of right cosets of $H$ in $G$ in such a
way that the $G$ action of $M$ corresponds to left multiplication on
$G/H$.  The derivative of $\phi_o$ at the identity $e \in G$ defines
a linear map
\begin{equation*}
  d\phi_o : \fg \to T_oM~,
\end{equation*}
where we have identified the tangent space $T_eG$ to the group at the
identity with the Lie algebra $\fg$.  Explicitly, if $X \in \fg$, then
\begin{equation*}
  d\phi_o(X) = \frac{d}{dt} \exp(tX) \cdot o \biggr|_{t=0}~.
\end{equation*}
This map is surjective with kernel the Lie subalgebra $\fh \subset
\fg$ corresponding to the isotropy subgroup $H$.  In other words, we
have an exact sequence:
\begin{equation}
  \label{eq:exact}
    \begin{CD}
    0 @>>> \fh @>>> \fg @>d\phi_o>> T_oM @>>> 0~.
  \end{CD}
\end{equation}
This is an exact sequence of $H$-modules, where $H$ acts on $\fh$
and $\fg$ via the adjoint representation and on $T_oM$ via the
\textbf{linear isotropy representation}; that is, if $h\in H$ and
$\boldsymbol{v}\in T_oM$, we define
\begin{equation}
  \label{eq:linisorep}
  h \cdot \boldsymbol{v} = \frac{d}{dt} h \cdot \gamma(t) \biggr|_{t=0}~,
\end{equation}
where $\gamma$ is a curve in $M$ through $o$ with tangent vector
$\boldsymbol{v}$.  This action makes $T_oM$ isomorphic to $\fg/\fh$ as
an $H$-module.

The metric $g$ defines an inner product $\left<-,-\right>$ on $T_oM$.
Invariance of $g$ under $G$ is equivalent to the invariance of
$\left<-,-\right>$ under $H$, whence the linear isotropy
representation defines a Lie algebra homomorphism $\fh \to
\fso(T_oM)$.  More generally, there is a bijective correspondence
between $H$-invariant tensors on $T_oM$ and $G$-invariant tensor
fields on $M$.

We can realise the linear isotropy representation explicitly by
choosing a complement $\fm$ of $\fh$ in $\fg$, so that $\fg = \fh
\oplus \fm$, and defining the action of $h\in H$ on $X \in \fm$ by
\begin{equation*}
  h \cdot X =  \left(\Ad(h) X\right)_\fm~,
\end{equation*}
where, here and in the following, the subscript $\fm$ indicates the
projection onto $\fm$ along $\fh$; that is, we simply discard the
$\fh$-component of $\Ad(h) X$.  If $\fm$ is stable under $\Ad(H)$, so
that the projection is superfluous, we say that $\fg = \fh\oplus \fm$
is a \textbf{reductive split}, and the pair $(\fg,\fh)$ is said to be
\textbf{reductive}.  This is equivalent to the splitting (in the sense
of homological algebra) of the exact sequence \eqref{eq:exact} in the
category of $H$-modules.

One often says that the manifold $(M,g)$ is ``reductive,'' but this is
an abuse of language.  It is important to stress that reductivity is
not an intrinsic geometric property of $(M,g)$ but of the linear
isotropy representation, whence of the description of $M$ as a coset
space $G/H$.  In fact, there are homogeneous spaces $(M,g)$ admitting
different coset descriptions $G_1/H_1$ and $G_2/H_2$, say, where one
of them is reductive but not the other.  The Kaigorodov space,
discussed in Section~\ref{sec:kaigorodov}, is one such example: it is
a left-invariant metric on a Lie group (whence trivially reductive),
but as a homogeneous space of its full isometry group, it is
nonreductive.  Nevertheless we will say that a homogeneous
pseudo-riemannian manifold $(M,g)$ is \textbf{reductive} if there
exists some $G$ acting transitively on $M$ via isometries, with
typical isotropy $H$ and for which the pair $(\fg,\fh)$ is reductive.

If $(M,g)$ is riemannian then $\Ad(H)$ is compact, whence we can
always find a reductive split.  This is done as follows: choose any
positive-definite inner product on $\fg$ and average over $\Ad(H)$ to
make it invariant.  Then let $\fm = \fh^\perp$ be the perpendicular
complement of $\fh$.  Since $\fh$ is a submodule, so is $\fm$.
If $g$ has indefinite signature, however, reductivity is a nonempty
condition.  Nevertheless, it appears that all four-dimensional
lorentzian homogeneous spaces \cite{Komrakov} are indeed reductive.

\subsection{Reductive homogeneous structures}
\label{sec:structures}

For the present purposes, the importance of reductivity stems from a
theorem of Ambrose and Singer \cite{AmbroseSingerHomog} in the
riemannian case and extended to the pseudo-riemannian case by Gadea
and Oubiña \cite{GadeaOubina}, which provides an alternate
characterisation of reductive (locally) homogeneous manifolds.  As
reformulated by Kostant \cite{KostantHomog}, the theorem states that
$(M,g)$ is a reductive locally homogeneous pseudo-riemannian manifold
if and only if there exists a metric linear connection with parallel
torsion and parallel curvature.  In other words, $(M,g)$ is reductive
locally homogeneous if and only if there exists a connection
$\widetilde\nabla$ on the tangent bundle, with torsion
\begin{equation*}
  \widetilde T(X,Y) = \widetilde\nabla_X Y - \widetilde\nabla_Y X -
  [X,Y]
\end{equation*}
and curvature
\begin{equation*}
  \widetilde R(X,Y) Z = \widetilde\nabla_{[X,Y]} Z - \widetilde\nabla_X
  \widetilde\nabla_Y Z +   \widetilde\nabla_Y \widetilde\nabla_X Z~,
\end{equation*}
and such that
\begin{equation}
  \label{eq:AmbroseSinger}
  \widetilde\nabla g = 0 \qquad
  \widetilde\nabla \widetilde T = 0\qquad
  \widetilde\nabla \widetilde R = 0~.
\end{equation}
These conditions define a nontrivial generalisation of the notion of a
locally symmetric space, where $\widetilde T = 0$ and hence
$\widetilde\nabla$ is the Levi--Cività connection.  In general, the
connection $\widetilde\nabla$ is called the \textbf{canonical
  connection}.

This connection can be equivalently characterised as the
$\fh$-component of the left-invariant Maurer--Cartan one-form
$\btheta$ on $G$, thought of as the total space of the principal
$H$-bundle $G \to G/H$.  Indeed, $\btheta$ is a $\fg$-valued one-form
on $G$.  Under the reductive split $\fg = \fh \oplus \fm$, we can
decompose $\btheta$ into a component along $\fh$ (the canonical
connection) and a component along $\fm$ (the vielbein).  Clearly,
$\fm$ is recovered as the kernel of the canonical connection. In other
words, given $\widetilde\nabla$ satisfying the conditions
\eqref{eq:AmbroseSinger}, one recovers the reductive split $\fh \oplus
\fm$ by declaring $\fm \subset \fg$ to correspond to those Killing
vectors which are $\widetilde\nabla$-parallel at the identity coset
$o$. The geodesics of the canonical connection are given by curves of
the form $\exp( t X ) \cdot o$ with $X \in \fm$, where $o \in M$ is
the identity coset.\footnote{For the rest of this paper, when
  unspecified, the word geodesic will be reserved for those of the
  Levi--Cività connection.}

The difference between the connection $\widetilde \nabla$ and the
Levi-Cività connection $\nabla$ is a $(2,1)$-tensor $S: TM \to \End
TM$, defined by
\begin{equation*}
  S_Y X := \nabla_X Y - \widetilde\nabla_X Y~.
\end{equation*}
In fact, since both connections preserve the metric,
\begin{equation*}
  g(S_X Y, Z) = - g(Y, S_X Z)~,
\end{equation*}
whence $S : TM \to \fso(TM)$.  The space of such tensors are sections
of a vector bundle $T^*M \otimes \fso(TM)$ associated to the bundle of
orthonormal frames.  Using the metric we can think of this
equivalently as the sub-bundle $\eT = T^*M \otimes \Lambda^2 T^*M
\subset \otimes^3 T^*M$.  This corresponds to thinking of $S$ as a
trilinear map
\begin{equation*}
  S(X,Y,Z) = g(S_X Y, Z)~.
\end{equation*}
In generic dimension (here, $\dim M > 2)$, the bundle $\eT$ breaks up
into the Whitney sum of three sub-bundles
\begin{equation*}
  \eT = \eT_1 \oplus \eT_2 \oplus \eT_3~,
\end{equation*}
each one corresponding to an irreducible representation of the
orthogonal group.  In terms of Young tableaux, this decomposition is
given by
\begin{equation*}
  \begin{aligned}[t]
    T^* \otimes \Lambda^2 T^*  &= \eT_1 \oplus \eT_2 \oplus \eT_3\\
    \yng(1) \otimes \yng(1,1) & = \yng(1) \oplus \yng(2,1) \oplus
    \yng(1,1,1)~.
  \end{aligned}
\end{equation*}
More explicitly, the bundles $\eT_i$ can be described as follows:
\begin{enumerate}
\item Sections of $\eT_1$ correspond to sections $S$ of $\eT$ such
  that
  \begin{equation*}
    S(X,Y,Z) = g(X,Y) \alpha(Z) - g(X,Z) \alpha(Y)
  \end{equation*}
  for some one-form $\alpha$, whence $\eT_1 \cong T^*M$.
\item Sections of $\eT_2$ correspond to sections $S$ of $\eT$ which
  satisfy
  \begin{equation}
    \label{eq:T1+T2}
    S(X,Y,Z) + S(Y,Z,X) + S(Z,X,Y) = 0~,
  \end{equation}
  and are in the kernel of the map $C: \otimes^3 T^*M \to T^*M$
  defined by contracting with the (inverse) metric on the first two
  indices:
  \begin{equation}
    \label{eq:cmap}
    C(S)(X) = \sum_{a,b} g^{ab} S(\be_a, \be_b, X)~,
  \end{equation}
  where $\be_a$ is a pseudo-orthonormal frame and $g^{ab}$ is the
  inverse of $g_{ab} = g(\be_a, \be_b)$.
\item Sections of $\eT_3$ correspond to sections of $\eT$ which are
  totally skew-symmetric, whence $\eT_3 \cong \Lambda^3 T^*M$.
\end{enumerate}

It is easy to write down the explicit expressions for each of the
components of $S$.  We will write $S_{abc} = S(\be_a,\be_b,\be_c)$
relative to a pseudo-orthonormal frame.  Then
\Yboxdim{3pt}
\begin{equation*}
  S_{abc} = S^{\,\yng(1)}_{abc} + S^{\,\yng(2,1)}_{abc} +
  S^{\,\yng(1,1,1)}_{abc}~,
\end{equation*}
where
\begin{align*}
  S^{\,\yng(1)}_{abc} &= g_{ab} \xi_c - g_{ac} \xi_b\\
  S^{\,\yng(1,1,1)}_{abc} &= \tfrac13 \left( S_{abc} + S_{bca} + S_{cab}
  \right)\\
  S^{\,\yng(2,1)}_{abc} &= S_{abc} - S^{\,\yng(1)}_{abc} -
  S^{\,\yng(1,1,1)}_{abc}~,
\end{align*}
where
\begin{equation*}
  \xi_c = \tfrac{1}{n-1} g^{ab} S_{abc}~,
\end{equation*}
with $n = \dim M$.

Given a reductive locally homogeneous space $(M,g)$, the
Ambrose--Singer theorem guarantees the existence of a tensor $S$,
called a \textbf{(reductive) homogeneous structure}.  Following
Tricerri and Vanhecke \cite{TricerriVanhecke}, we can distinguish
eight ($=2^3$) types of homogeneous structures, depending on whether
$S$ does or does not have a component in each of the three irreducible
components $\eT_i$.  This condition can be probed at any one point
$o\in M$ because $S$ is parallel with respect to a metric connection
$\widetilde\nabla$, hence its type under the orthogonal group does not
change under parallel transport with respect to $\widetilde\nabla$.

The eight possible homogeneous structures are the
following:\footnote{We abuse notation slightly and identify the
  bundles $\eT_i$ with their sheaves of sections, whence $S \in \eT_i$
  means that $S$ is a section of $\eT_i$, etc}
\begin{enumerate}
\item $S=0$: the locally symmetric spaces;
\item $S \in \eT_1$: here there exists a vector $\xi$ such
  that $S$ takes the form
  \begin{equation*}
    S_X Y = g(X,Y) \xi - g(Y,\xi) X~.
  \end{equation*}
  In riemannian signature, Tricerri and Vanhecke
  \cite{TricerriVanhecke} proved that $(M,g)$ has constant negative
  curvature, whence it is locally isometric to hyperbolic space; that
  is, locally symmetric.    In lorentzian signature, we must
  distinguish between two cases: according to whether the norm of
  $\xi$ is zero or nonzero.  In the latter case, Gadea and Oubiña
  \cite{GadeaOubina} proved that $(M,g)$ is locally
  isometric to anti-de~Sitter space, whereas if $\xi$ is null, then
  Montesinos Amilibia \cite{MontesinosAmilibia} showed that $(M,g)$ is
  a singular homogeneous plane-wave \cite{BOLhpw}:
  \begin{equation*}
    g = 2 du dv + A(\bx,\bx) \frac{du^2}{u^2} + |d\bx|^2~,
  \end{equation*}
  with $A$ a constant bilinear form;
\item $S \in \eT_2$;
\item $S \in \eT_3$: these are the so-called naturally
  reductive homogeneous spaces.  They can be alternatively
  characterised as those reductive homogeneous manifolds for which
  $\exp(t X) \cdot o$ is a geodesic through $o$ for every nonzero $X
  \in \fm$;
\item $S \in \eT_1 \oplus \eT_2$: in this case the tensor
  $S$ satisfies equation \eqref{eq:T1+T2};
\item $S \in \eT_2 \oplus \eT_3$: in this case the tensor
  $S$ is in the kernel of the contraction map $C$ defined in
  \eqref{eq:cmap};
\item $S \in \eT_1 \oplus \eT_3$: in this case, the tensor
  $S$ satisfies
  \begin{equation*}
    S(X,Y,Z) + S(Y,X,Z) = 2 g(X,Y) \alpha(Z) - g(X,Z) \alpha(Y) -
    g(Z,Y) \alpha(X)~,
  \end{equation*}
  for some one-form $\alpha$.  It is shown in \cite{PatrickT1T3} that
  if $\alpha$ has non-zero norm, then the underlying geometry is once
  again that of a symmetric space, whereas if $\alpha$ is null then it
  is a generic singular homogeneous plane-wave \cite{BOLhpw}
  \begin{equation*}
    g = 2 du dv + \left[ A(e^{-u F}\bx,e^{-u F}\bx) +
      2v\right] du^2 + |d\bx|^2~,
  \end{equation*}
  where $A$ is once again a constant bilinear form and $F$ is a
  skew-symmetric matrix; and, finally,
\item $S$ generic.
\end{enumerate}

It must be stressed that a given homogeneous space can admit more than
one homogeneous structure, as indeed can be seen from the examples in
Section~\ref{sec:examples} or from the characterisation of the
non-degenerate $\eT_{1}$ class.  We can understand this as follows.
There is a one-to-one correspondence between homogeneous structures
$S$ and reductive splits $\fg = \fh \oplus \fm$.  In principle,
different choices of $\fh$ and $\fm$ give rise to different
homogeneous structures.  Indeed, given $\fg = \fh \oplus \fm$, the
homogeneous structure $S$ at the identity coset $o$ is given by
\begin{equation}
  \label{eq:S@o}
  S(X,Y,Z) = g \left( \nabla_Y X \bigr|_o, Z \right)~,
\end{equation}
where $X,Y,Z$ are Killing vectors in $\fm$.

Now suppose that $\fg = \fh \oplus \fm$ is a reductive split with
maximal $\fg$, and let $\fg' \subseteq \fg$ be a subalgebra such that
the restriction of the map $\fg \to T_o M$ to $\fg'$ is still
surjective.  Let $\fh' = \fg' \cap \fh$ and let $\fm' = \fg' \cap
\fm$.  Surjectivity implies that $\fm' = \fm$, whence $\fg' = \fh'
\oplus \fm$ is still a reductive split.  This split can be deformed as
follows.  We pick a subspace $\fm' \subset \fg'$ such that $\fg' =
\fh' \oplus \fm'$ is still a reductive split.  This means that $\fm'$
is the graph of an $\fh'$-equivariant linear map $\varphi: \fm \to
\fh'$; that is,
\begin{equation*}
  \fm' = \left\{ \varphi(X) + X \mid X \in \fm  \right\}~.
\end{equation*}
As $\fh'$ grows, there are more linear maps $\fm \to \fh'$, but
also the $\fh'$-equivariance condition becomes stronger.  It is
therefore not inconceivable that we should obtain nontrivial
$\varphi$'s by restricting to subalgebras as just described.  We
will see an example of this in Section~\ref{sec:examples}.
Observe, by the way, that conjugate subalgebras yield isomorphic
homogeneous structures.

Given a homogeneous structure $S$, the Lie bracket restricted to
the subspace $\fm$ of the isometry algebra is given by the formula
\begin{equation}
  \label{eq:Salgebra}
  [X,Y] = S_X Y - S_Y X + \widetilde{R}(X,Y)~,
\end{equation}
where $X,Y\in\fm$ and $S$ and $\widetilde{R}$ are evaluated at the
point $o$. This defines the subspace $\fm \oplus [\fm,\fm]$, from
which we may define the full reductive split $\fm \oplus \fh$ to
be the algebraic closure of this subspace under the Lie bracket
\eqref{eq:Salgebra} together with
\begin{equation}
[A,X]=A(X) \quad \textrm{and} \quad [A,B] = AB - BA ~,
\end{equation} where $X \in \fm$ and $ A,B \in \End (\fm)$. Notice
that not all elements of $\fh$ need appear in $\widetilde{R}$, in
fact the holonomy algebra $\text{hol}(\widetilde{\nabla})$ must be an
ideal of $\fh$.

\subsection{Calculating with homogeneous spaces}
\label{sec:calc}

We now collect some useful formulae for calculating the Riemann tensor
of a homogeneous space in terms of Lie algebraic data.  For more
details one can consult, for example, the book \cite{Besse}.

Let $X,Y,Z$ be Killing vectors on $M = G/H$.  The Koszul formula for
the Levi-Cività connection reads
\begin{equation}
  \label{eq:Koszul}
  g(\nabla_X Y, Z) = \half g([X,Y],Z) + \half g([X,Z],Y) + \half
  g(X,[Y,Z])~.
\end{equation}
At the identity coset $o\in M$ and assuming that $X,Y,Z$ are Killing
vectors in $\fm$, then
\begin{equation}
  \label{eq:levicivita}
  \nabla_X Y\bigr|_o = -\half [X,Y]_{\fm} + U(X,Y)~,
\end{equation}
where $U:\fm \times \fm \to \fm$ is a symmetric tensor given
by\footnote{The apparent difference in sign between equation
  \eqref{eq:Koszul} and equations \eqref{eq:levicivita} and
  \eqref{eq:Utensor} stems from the fact that Killing vectors on $G/H$
  generate left translations on $G$, whence they are right-invariant.
  Thus the map $\fg \to \text{Killing vector fields}$ is an
  anti-homomorphism.}
\begin{equation}
  \label{eq:Utensor}
  \left<U(X,Y),Z\right> = \half \left<[Z,X]_{\fm},Y\right> +
  \half \left<[Z,Y]_{\fm},X\right>~,
\end{equation}
for all $Z\in\fm$.  It should be remarked that \eqref{eq:levicivita}
is only valid at $o\in M$, since $\nabla_X Y$ is not generally a
Killing vector. Of course, since $\nabla$ is G-invariant, then one can
determine $\nabla_X Y\bigr|_p$ at any other point by acting with any
isometry relating $o$ and $p$.

The formula \eqref{eq:S@o} for the corresponding homogeneous structure
(at $o$) can now be written explicitly:
\begin{equation}
  \label{eq:S@oexplicit}
  S(X,Y,Z) = \half \left< [X,Y]_\fm, Z \right> + \half \left<
    [Z,X]_\fm, Y \right> +\half \left< [Z,Y]_\fm, X \right>~,
\end{equation}
for $X,Y,Z \in \fm$.

The $U$-tensor is not generally invariant under the linear isotropy
representation; indeed, for all $Z\in\fh$,
\begin{equation*}
  (Z\cdot U)(X,Y) = [[Z,X]_{\fh}, Y]_{\fm} + [[Z,Y]_{\fh}, X]_{\fm}~;
\end{equation*}
although it clearly does when $G/H$ is reductive.  The vanishing of
the $U$-tensor characterises the \textbf{naturally reductive}
homogeneous structures.

The Riemann curvature tensor is $G$-invariant and it can be computed
at $o$.  One obtains, for $X,Y,Z,W$ vectors in $\fm$, the curvature
tensor at $o$ is given by
\begin{multline}
  \label{eq:riemann}
  R(X,Y,Z,W) = \left<U(X,W),U(Y,Z)\right>
  - \left<U(X,Z),U(Y,W)\right>\\
  + \tfrac1{12} \left<[X,[Y,Z]]_{\fm},W\right>
  - \tfrac1{12} \left<[X,[Y,W]]_{\fm},Z\right>\\
  - \tfrac16 \left<[X,[Z,W]]_{\fm},Y\right>
  - \tfrac1{12} \left<[Y,[X,Z]]_{\fm},W\right>\\
  + \tfrac1{12} \left<[Y,[X,W]]_{\fm},Z\right>
  + \tfrac16 \left<[Y,[Z,W]]_{\fm},X\right>\\
  - \tfrac16 \left<[Z,[X,Y]]_{\fm},W\right>
  - \tfrac1{12} \left<[Z,[X,W]]_{\fm},Y\right>\\
  + \tfrac1{12} \left<[Z,[Y,W]]_{\fm},X\right>
  + \tfrac16 \left<[W,[X,Y]]_{\fm},Z\right>\\
  + \tfrac1{12} \left<[W,[X,Z]]_{\fm},Y\right>
  - \tfrac1{12} \left<[W,[Y,Z]]_{\fm},X\right>\\
  - \half \left<[X,Y]_{\fm},[Z,W]_{\fm}\right>
  - \tfrac14 \left<[X,Z]_{\fm},[Y,W]_{\fm}\right>
  + \tfrac14 \left<[X,W]_{\fm},[Y,Z]_{\fm}\right>~,
\end{multline}
which can be obtained by polarisation from the simpler expression for
$K(X,Y):=\left<R(X,Y)X,Y\right>$, which is also easier to derive.
Indeed, and for completeness, one has
\begin{multline*}
  6 R(X,Y,Z,W) = K(X+Z,Y+W) - K(Y+Z,X+W)\\
  - K(Y+W,X) + K(Y+Z,X) - K(X+Z,Y) + K(X+W,Y)\\
  - K(Y+W,Z) + K(X+W,Z) - K(X+Z,W) + K(Y+Z,W)\\
  + K(X,W) - K(X,W) - K(Y,W) + K(Y,Z) - K(X,Z)~,
\end{multline*}
where
\begin{multline*}
  K(X,Y) = - \tfrac34 |[X,Y]_{\fm}|^2 - \half
  \left<[X,[X,Y]]_{\fm},Y\right> - \half
  \left<[Y,[Y,X]]_{\fm},X\right>\\
  + |U(X,Y)|^2 - \left<U(X,X),U(Y,Y)\right>
\end{multline*}
and where $|-|^2$ is the (indefinite) norm associated to
$\left<-,-\right>$.

\subsection{Geodesics in homogeneous spaces}
\label{sec:geodesics}

We shall be interested in geodesics in $G/H$ which are themselves
orbits of one-parameter subgroups of $G$.  By homogeneity we can
assume that the geodesics pass through our base point $o$.  Such
\textbf{homogeneous} geodesics can always be reparameterised so that
they are given by
\begin{equation}
  \label{eq:homogeod}
  \gamma(t) = \exp( t X ) \cdot o~,
\end{equation}
for some \textbf{geodetic vector} $X \in \fg$.  The condition of
$X\in\fg$ being geodetic is that the curve traced by $\gamma$ above be
a geodesic.  If $\gamma'$ has non-zero norm, this is equivalent to
the self-parallel condition $\nabla_{\gamma'} \gamma' = 0$, but
for null geodesics, one can relax this condition to
$\nabla_{\gamma'} \gamma' = c\left(\gamma\right) \gamma'$.  It follows from the
Koszul formula~\eqref{eq:Koszul} that $X\in\fg$ is geodetic if and
only if
\begin{equation}
  \label{eq:geodetic}
  \left<[X,Z]_{\fm},X_\fm\right> = c \left<X_\fm, Z\right>~,
\end{equation}
for all $Z \in \fm$, and where $c$ is some constant.

If $X$ in equation~\eqref{eq:homogeod} belongs to $\fm$, then we say
that the geodesic is \textbf{canonically homogeneous}, since then
$\gamma$ is also a geodesic for the canonical connection.

There are some spaces for which all geodesics are homogeneous.  This
is the case, for example, for the naturally reductive spaces in which
the $U$ tensor defined in \eqref{eq:Utensor} vanishes.  In fact, for
such spaces every geodesic is canonically homogeneous.  More
generally, a space for which all geodesics are homogeneous is called a
\textbf{geodesic orbit} space or \textbf{g.o.}~space, for short.  In
Section~\ref{sec:kaplan} we will discuss an example of a g.o.~space, a
six-dimensional lorentzian manifold of the type first considered by
Kaplan (see, for example, \cite{KowNik}).  In a g.o.~space, given any
nonzero $X\in\fm$, there is an element $\phi(X)\in\fh$ such that $X +
\phi(X) \in \fg$ is geodetic; that is,
\begin{equation}
  \label{eq:graph}
  \left<[\phi(X),X]_{\fm} + c(X) X, Z\right> =
  \left<[X,Z]_{\fm},X\right>~,
\end{equation}
for all $Z\in\fm$.

\subsection{Homogeneous plane waves}
\label{sec:HSwaves}

Homogeneous plane waves have been recently classified.  In
\cite{BOLhpw} it is proved that they fall in two classes: one
consisting of regular homogeneous plane waves and one consisting of
singular homogeneous plane waves.

In Brinkmann coordinates the regular plane-wave metric takes the form
\begin{equation}
  \label{eq:reg_hom_waves}
  2 du dv + A(e^{-uF} \bx, e^{-uF} \bx) du^2 + |d\bx|^2~,
\end{equation}
where $A$ is a constant symmetric bilinear form and $F$ is a constant
skew-symmetric matrix.  When $[F,A]=0$, $F$ drops out of
the metric and the resulting metric is symmetric.  Otherwise, the
space is not locally symmetric, but is a naturally reductive space as
evidenced by the existence of a $\eT_3$ structure
\begin{equation}
  \label{eq:reg_hom_wave_hom_struct}
  S = \tfrac12 F_{ij} du \wedge dx^i \wedge dx^j~.
\end{equation}
Furthermore, it can be shown that regular homogeneous plane waves
do not admit homogeneous structures of type $\eT_1\oplus\eT_3$.

The metric for the singular homogeneous plane waves in Brinkmann
coordinates reads
\begin{equation*}
  2 e^z dz dv + A( e^{-z F}\bx, e^{-z F}\bx) dz^2 +  |d\bx|^2~.
\end{equation*}
We can change coordinates and rewrite this metric as
\begin{equation}
  \label{eq:sing_hom_waves}
  2 du dv + A( e^{-(\log u)F}\bx, e^{-(\log u)F}\bx)
    \frac{du^2}{u^2} + |d\bx|^2~,
\end{equation}
which has manifestly a pp-wave singularity at $u=0$, whence it is
incomplete.  In general it admits a $\eT_1\oplus \eT_3$ structure
given by
\begin{equation}
  \label{eq:sing_hom_wave_hom_struct}
  S_{uuv} = \frac{1}{u} \qquad S_{uij} = \frac{1}{u}F_{ij} \qquad
  S_{iuj} = \frac{1}{u}\left[ \delta_{ij}  - F_{ij} \right]~,
\end{equation}
and it admits an $\eT_1$ structure when $[F,A] =0$, which is just the
same as taking $F=0$.  In fact, as was said in
Section~\ref{sec:structures}, the singular homogeneous plane waves
with $F=0$ are the only spacetimes admitting a degenerate $\eT_1$
structure.

Let us remark that if one tries to repeat the analysis of
\cite{BOLhpw} by means of the Ambrose--Singer equations based on the
metric
\begin{equation*}
    2 du dv + A_{ij}(u) x^i x^j du^2 + |d\bx|^2~,
\end{equation*}
one finds that the only solutions for $S_{uuv}$ are either zero or
$1/(u+u_{0})$, signalling a regular or a singular plane wave,
respectively.

\section{Homogeneous plane-wave limits}
\label{sec:homogPL}

In this section we describe in very concrete terms the plane-wave
limit of a homogeneous spacetime along a homogeneous null geodesic.
We give two derivations of algebraic formulae---equations
\eqref{eq:Fab} and \eqref{eq:Aab}---for the limiting metric in terms
of the initial data describing the homogeneous spacetime and the
geodesic in question.  One derivation uses the covariant
characterisation of the plane-wave limit given in \cite{CovariantPL},
whereas the other involves a limiting procedure different yet
equivalent, as we will show, to the plane-wave limit, and which does
away with the need to to find neither an adapted frame nor adapted
coordinates.  We start, though, with a brief review of the plane-wave
limit itself.

\subsection{Plane-wave limits}
\label{sec:planewavelimits}

Let $\gamma$ be a null geodesic in a spacetime $(M,g)$. Then according
to Penrose \cite{PenrosePlaneWave} a coordinate system $(u,v,y^i)$ can
be found in a neighbourhood of $\gamma$ without conjugate points, such
that the metric takes the form
\begin{equation}
  \label{eq:twistfree}
  g = du dv + \alpha dv^2 + \sum_i \beta_i dy^i dv +
  \sum_{i,j} C_{ij} dy^i dy^j~,
\end{equation}
where $\alpha, \beta_i$ and $C_{ij}$ are smooth functions and we have
$\gamma' =\partial_{u}$.  Following Penrose, we now rescale the
coordinates by a factor $\Omega \in \RR$
\begin{equation*}
  v \mapsto \Omega^2 v, \quad y^i \mapsto \Omega y^i~, \quad u
  \mapsto u~, \end{equation*}
and denote the metric we obtain from this redefinition by
$g_{\Omega}$.  Then the limit as $\Omega \to 0$ of $\Omega^{-2}
g_{\Omega}$ is well defined
\begin{equation*}
  \begin{split}
    \overline g & = \lim_{\Omega \to 0} \left( du dv + \Omega^2 \alpha
      dv^2 + \Omega\sum_i \beta_i dy^i dv + \sum_{i,j} C_{ij}
    dy^i dy^j \right) \\
    & = du dv + \sum_{i,j} C_{ij}(u,0,0) dy^i dy^j
  \end{split}
\end{equation*}
and is called the \textbf{Penrose} or \textbf{plane-wave limit along
  $\gamma$}.  This limiting procedure has been extended by Güven
\cite{GuevenPlaneWave} to supergravity theories with additional fields
and provides a method for obtaining new supergravity solutions from
old.

As was shown in \cite{Limits}, there are geometric properties of
the spacetime which are preserved under plane-wave limits.
Following Geroch \cite{Geroch} one calls these properties
\textbf{hereditary}, since the limiting spacetime inherits them
from the parent spacetime. Such hereditary properties include the
Killing spinors and Killing vectors; although it is not uncommon
that the limiting spacetime is more (super)symmetric than the
parent spacetime.

Another example of a hereditary property \cite{Geroch,Limits} is that
of being locally symmetric; that is, if the Riemann curvature tensor
is parallel before the limit, it is parallel after the limit.
Furthermore, since every locally symmetric plane-wave, or
Cahen--Wallach space, is geodesically complete, we can strengthen this
result and claim that the plane-wave limit of a locally symmetric
space is symmetric, after suitable completion.  Since a locally
symmetric space is locally homogeneous, we see that homogeneity can be
hereditary under the plane-wave limit even though this is not the case
in general, as evidenced by the Kaigorodov space \cite{Patricot}.
This prompts the following question: under what extra conditions is
(local) homogeneity preserved?  The remainder of this section is
devoted to exploring this question.

If $\gamma$ is a homogeneous geodesic then it was shown in
\cite{SimonPL} that the plane-wave limit along $\gamma$ is locally
homogeneous as defined in Section~\ref{sec:basic}.  This was done
by examining the Killing transport along $\gamma$ and showing that
it has a well-defined plane-wave limit which generates a Killing
vector field for the plane-wave limit metric. As the generic plane
wave is of cohomogeneity one, this extra Killing vector gives the
result.

The above gives a sufficient condition on a null geodesic, in a
generic spacetime, for the plane-wave limit along it to be
homogeneous. It is however not a necessary condition as the
following example shows.  Consider the metric
\begin{equation*}
  2du dv + u dv^2 + \sqrt{u} \sum_i (dx^i)^2~.
\end{equation*}
This is an incomplete and nonhomogeneous metric, with no Killing
vector in the $\partial_u$ direction.  Therefore the null geodesic
given by $\partial_u$ is not homogeneous.  The plane-wave limit
along this geodesic,
\begin{equation*}
  2du dv + \sqrt{u} \sum_i (dx^i)^2~,
\end{equation*}
is however a singular homogeneous plane wave \cite{PRTwaves}.

The above raises the question about the existence of homogeneous
(null) geodesics in reductive homogeneous spaces.  It is a theorem by
Kowalski and Szenthe \cite{KowSze}, suitably extended to the
lorentzian case \cite{SimonPL}, that in a reductive lorentzian
homogeneous manifold there exists at least one (not necessarily null)
homogeneous geodesic through every point.  On the other hand, on a
g.o.~spacetime every geodesic is homogeneous, whence homogeneity is
hereditary for these spacetimes.

The existence of particular classes of homogeneous structures on a
reductive homogeneous spacetime can indicate the existence of null
homogeneous geodesics. Indeed, if $S$ is a section of $\eT_1
\oplus \eT_3$, then for a null geodesic $\gamma$ of the
$\widetilde{\nabla}$ connection we have
\begin{equation*}
  0 = \widetilde{\nabla}_{\gamma'} \gamma' = \nabla_{\gamma'} \gamma'-
  g(\gamma' , \gamma') \xi + g(\gamma' , \xi ) \gamma' =
  \nabla_{\gamma'} \gamma' +g(\gamma', \xi) \gamma'~.
\end{equation*}
Now if we reparameterise $\gamma(\tau)$ to $\overline\gamma(s)$,
such that $\gamma' =\partial_{\tau} = f(s) \partial_s =
f(s)\overline\gamma'$, we find that
\begin{equation*}
  0 = f^2 \nabla_{\overline\gamma'} \overline\gamma'
    + f \left(\nabla_{\overline\gamma'} f \right) \overline\gamma'
    + f^2 g(\overline\gamma'\ ,\xi ) \overline\gamma' \; .
\end{equation*}
So that a solution to
\begin{equation*}
  \frac{\partial f}{\partial s} + g(\overline\gamma'\ ,\xi ) f =0
\end{equation*}
maps a null geodesic of $\widetilde{\nabla}$ to a null geodesic of
$\nabla$.  Conversely, given a null geodesic for $\nabla$ we can
perform the inverse transformation and obtain a null geodesic for
$\widetilde{\nabla}$.  Thus, every null geodesic in a spacetime with a
homogeneous structure of type $\eT_1 \oplus \eT_3$ is canonically
homogeneous and hence the plane-wave limit of the spacetime is always
a homogeneous plane wave admitting a homogeneous structure contained
in $\eT_{1}\oplus\eT_{3}$.  In fact, it has been shown in
\cite{PatrickT1T3} that a reductive homogeneous space admitting a
structure of type $\eT_{1}\oplus\eT_{3}$ is either a locally symmetric
space (and hence naturally reductive), a singular homogeneous
plane-wave or a more general naturally reductive space, depending on
whether the vector in the $\eT_1$-component is nondegenerate, null or
zero, respectively.

In Section~\ref{sec:homstructpl} we will show, by considering various
examples, that the existence of one of the other classes of
homogeneous structures on a spacetime says little about the existence
of homogeneous geodesics.

\subsection{The covariant method}

Let $g$ be a lorentzian metric and $\gamma$ a null geodesic of
$g$. Consider $g$ to be written in a twist-free coordinate system
\eqref{eq:twistfree} and let $(\partial_u, \partial_v, \partial_i)$
denote the dual frame to $(du, dv, dy^i)$.

In \cite{CovariantPL}, the following covariant formulation of the
plane-wave limit is given.  We say that a local frame
$(E_+,E_-,E_a)$ is \textbf{adapted to a null geodesic $\gamma$},
if the following conditions are satisfied:
\begin{enumerate}
\item $E_+$ is a geodesic vector field such that $E_+|_{\gamma}$ is
  proportional to $\partial_u |_{\gamma}$, where $u$ is the parameter
  along $\gamma$;
\item $\nabla_u E_- = \nabla_u E_a = 0$ along $\gamma$; and
\item the metric takes the form
  \begin{equation*}
    g = 2 \theta^+ \theta^- + \sum_a \theta^a \theta^a
  \end{equation*}
  where the $\theta$'s are the dual coframe.
\end{enumerate}

Let $(E_+,E_-,E_a)$ be such an adapted frame.  We can write $E_a$ in
the form
\begin{equation*}
  E_a = E_a^i \partial_i + E_a^u \partial_u + E_a^v \partial_v~.
\end{equation*}
By taking its inner product with $E_+$ and with $E_b$ we see that,
restricted to the geodesic $\gamma$, we have
\begin{equation*}
  E_a^v =0
\end{equation*}
and
\begin{equation*}
  E_{ai}E_b^i = C_{ij} E_a^j E_b^i = \delta_{ab}~.
\end{equation*}
Calculating the covariant derivative of $E_a$ we have
\begin{equation*}
   (E_a^i)' + E_a^j \Gamma_{ju}^i = 0
\end{equation*}
and the dual equation
\begin{equation*}
   (E_{ai})' - E_{aj} \Gamma_{iu}^j = 0~.
\end{equation*}
Thus
\begin{equation}
  \label{eq:BrinkSymm}
   (E_{ai})' E^i_b = -E_{ai} ( E_b^i)' = E_{ai} E_b^j \Gamma_{ju}^i
  = E_a^i E_{bj} \Gamma_{ju}^i = E_a^i ( E_{bj})'~.
\end{equation}

Now consider the plane-wave limit $\overline g$ of the metric $g$.
A frame $E_M$ satisfying equation \eqref{eq:BrinkSymm} defines a
change of coordinates from the Rosen coordinate description of
$\overline g$ to a Brinkmann coordinate description
\begin{equation*}
  2dx^+ dx^- + A_{ij}(x^+) x^i x^j (dx^+)^2 + \sum_i (dx^i)^2,
\end{equation*}
where
\begin{equation*}
  A_{ab}(x^+) = -R(E_+, E_a, E_+, E_b)|_{\gamma} = -R(E_+, \partial_i,
  E_+, \partial_j)|_{\gamma} E^i_a|_{\gamma} E^j_b|_{\gamma}~.
\end{equation*}

This covariant description of the plane-wave limit illustrates
that the limit is really an invariant of the null geodesic and not
just a remnant of a special coordinate system. However, it is not
much easier to apply than the usual plane-wave limit as finding a
parallel frame can be difficult. On the other hand, on reductive
spaces it is a fruitful approach.

Indeed, suppose that now $(M,g)$ is a reductive homogeneous space with
a homogeneous structure $S$. Let $M$ be locally isomorphic to the
quotient $G/H$ and let $\fg = \fm \oplus \fh$ be the reductive split
of the Lie algebra of $G$ associated to $S$. Let $U \in \fg$ be the
geodetic vector that determines $\gamma$ as homogeneous.  Let $V \in
\fm$ be the dual null vector and complete to a basis with orthonormal
elements $Y_i \in \fm$. The classification of homogeneous plane waves
\cite{BOLhpw} states that the plane-wave limit in Brinkmann
coordinates will be of the form:
\begin{equation*}
   A(x^+) = e^{x^+F} A_0 e^{-x^+F} \quad \textrm{or}
   \quad  A(x^+) = e^{\log (cx^+) F} A_0
   e^{-\log(cx^+)F}/(cx^+)^2 ~,
\end{equation*}
where $A_0$ is a nondegenerate symmetric bilinear form, $F$ is a
skew-symmetric bilinear form and $c \neq 0$ is the constant in
\eqref{eq:geodetic}. The first case corresponds to the
non-singular plane-waves and the second to the singular waves. We
shall take the origin $o$ for the non-singular waves to be the
point $(0,0,0)$, while for the singular waves we take $(1/c,0,0)$.

We will now use the above covariant description and the algebraic
description of the curvature tensor on such a background to write
down an algebraic formula for both $A_0$ and $F$.

Let $E_M$ be an adapted frame to the geodesic $\gamma$ which when
restricted to $o$ corresponds to the basis $(U,V,Y_i)$. For a
non-singular homogeneous plane-wave limit $\overline g$ we have
\begin{equation}
  \label{A(x^+)}
  \exp( x^+ [F, -]) \cdot A_0 = A_{ab}(x^+) = -R(E_+, E_a, E_+,
  E_b)|_{\gamma}~.
\end{equation}
Thus, evaluating at $o$,
\begin{equation} \label{A_0}
  \begin{split}
    (A_0)_{ab} &= -R(E_+, E_a, E_+, E_b)|_0 \\
    &= -R(U_{\fm}, Y_a, U_{\fm}, Y_b)~,
  \end{split}
\end{equation}
where $U_{\fm}$ is the projection to $\fm$ of $U \in \mathfrak{g}$
and $Y_a = E_a(0) \in \fm$. Similarly, we find that \eqref{A_0}
holds for the singular plane-waves.

Now, if we differentiate the left hand side of equation
\eqref{A(x^+)} and evaluate at $o$ we obtain
\begin{equation*}
    \frac{ \partial}{\partial x^+} \left(A_{ab}(x^+)
    \right)\Bigr|_o
    = -2 cA_0 + [F,A_0]~.
\end{equation*}
Differentiating the right hand side,
\begin{equation*}
  \begin{split}
    \frac{ \partial}{\partial x^+} \left(A_{ab}(x^+) \right)|_o &=
    -\frac{ \partial}{\partial x^+} R(E_+,E_a, E_+, E_b)|_o, \\
    &= -\nabla_{U} (R(E_+,E_a, E_+, E_b)|_{\gamma}), \\
    &=-\left(\nabla_{U} R(E_+,E_a, E_+, E_b) \right) |_{\gamma}, \\
    &=-\left(\nabla_{U} R\right)(E_+,E_a, E_+, E_b)  |_{\gamma},
  \end{split}
\end{equation*}
where we have used the fact that $U$ is a vector field tangent to
$\gamma$ and that the frame $E_M$ is parallel with respect to $U$.

The object $\nabla R$ is tensorial, that is
\begin{equation*}
  (\nabla R)(\cdot, \dots, fX, \dots, \cdot) = f(\nabla R)(\cdot,
  \dots,X,\dots, \cdot)~,
\end{equation*}
for any $f \in C^{\infty}(M)$, whence, by passing the restriction
to $0$ through the curvature, we have
\begin{equation*}
  \frac{ \partial}{\partial x^+} \left(A_{ab}(x^+) \right)|_0 =
  -\left(\nabla_{U_{\fm}} R\right)(U_{\fm},Y_a, U_{\fm}, Y_b)~.
\end{equation*}
As $U_{\fm}$ is a Killing vector \cite{Nomizu}
\begin{equation*}
  (\nabla_{U_{\fm}} - S_{U_{\fm}} \cdot)R = \mathcal{L}_{U_{\fm}} R =
  0~.
\end{equation*}
Hence we can replace the differential action of the covariant
derivative with the algebraic action of the linear map $S_{U_{\fm}}$,
\begin{equation*}
  \begin{split}
    \frac{ \partial}{\partial x^+} \left(A_{ab}(x^+) \right)|_0 &=
    -\left(S_{U_{\fm}} \cdot R\right)(U_{\fm},Y_a, U_{\fm}, Y_b), \\
    &=  R(S_{U_{\fm}} U_{\fm},Y_a, U_{\fm}, Y_b) + R(
    U_{\fm},S_{U_{\fm}}Y_a, U_{\fm}, Y_b) \\
    & \qquad + R( U_{\fm},Y_a, S_{U_{\fm}} U_{\fm}, Y_b) + R(
    U_{\fm},Y_a, U_{\fm}, S_{U_{\fm}}Y_b)~,
  \end{split}
\end{equation*}
where we have used that the action of $S_{U_{\fm}}$ annihilates
functions.  Therefore we obtain the formula
\begin{equation}
  \label{eq:[F,A_0]}
  \begin{split}
   -2c (A_0)_{ab}+ [F,A_0]_{ab} &=  R(S_{U_{\fm}} U_{\fm},Y_a, U_{\fm}, Y_b)+ R(
    U_{\fm},S_{U_{\fm}}Y_a, U_{\fm}, Y_b) \\
    &+ R( U_{\fm},Y_a,S_{U_{\fm}} U_{\fm} , Y_b)+  R( U_{\fm},Y_a,
    U_{\fm}, S_{U_{\fm}}Y_b)
  \end{split}
\end{equation}

Similarly, differentiating a second time and evaluating at zero,
we find that $(6c^2A_0-3c[F,A_0]+[F,[F,A_0]])_{ab}$ is given by
\begin{align*}
    &  R(S_{U_{\fm}} S_{U_{\fm}} U_{\fm},Y_a,
    U_{\fm}, Y_b)+ R( U_{\fm},S_{U_{\fm}}S_{U_{\fm}}Y_a, U_{\fm},
    Y_b)\\
    &+ R( U_{\fm},Y_a,S_{U_{\fm}}S_{U_{\fm}} U_{\fm},
    Y_b)
    + R(U_{\fm},Y_a, U_{\fm} , S_{U_{\fm}}S_{U_{\fm}}Y_b)\\
    &+ 2R(S_{U_{\fm}} U_{\fm},S_{U_{\fm}} Y_a, U_{\fm}, Y_b)
    + 2R(S_{U_{\fm}} U_{\fm},Y_a, S_{U_{\fm}} U_{\fm}, Y_b) \\
    &+
    2R(S_{U_{\fm}} U_{\fm},Y_a, U_{\fm}, S_{U_{\fm}}Y_b)
    + 2R( U_{\fm},S_{U_{\fm}}Y_a, S_{U_{\fm}}U_{\fm}, Y_b)\\
     &+
    2R(U_{\fm},S_{U_{\fm}}Y_a, U_{\fm}, S_{U_{\fm}} Y_b)
    + 2R(
    U_{\fm},Y_a, S_{U_{\fm}} U_{\fm} , S_{U_{\fm}}Y_b)\\
    &+ R(S_{S_{U_{\fm}}U_{\fm}} U_{\fm},Y_a, U_{\fm}, Y_b) +
    R(U_{\fm},S_{S_{U_{\fm}}U_{\fm}}Y_a, U_{\fm}, Y_b) \\
    &+
    R( U_{\fm},Y_a, S_{S_{U_{\fm}}U_{\fm}} U_{\fm},
    Y_b) + R(U_{\fm},Y_a, U_{\fm}, S_{S_{U_{\fm}}U_{\fm}}Y_b )~.
\end{align*}

Similar expressions can be obtained for higher order brackets
between $F$ and $A_0$.  By calculating enough terms of the form
$[F, \dots, [F,A_0]]$, one can solve for the skew-symmetric matrix
$F$, but in fact, it is not difficult to write down a general
solution.

First we note that since $U$ is geodetic, we have
\begin{equation*}
   S_{U_{\fm}} U_{\fm} +S_{U_{\fh}} U_{\fm}=S_{U} U_{\fm} = -c
   ~U_{\fm}~,
\end{equation*}
where we are extending\footnote{This is clearly consistent with its
  definition on $\fm$, as the canonical connection vanishes there.  In
  this way it denotes the skew-symmetric endomorphism $-A_X$ of $TM$
  associated to a Killing vector, as described, for example, in
  \cite{FMPHom}.  Notice, though, that strictly speaking this is an
  abuse of notation since $S$ is tensorial, so that $S(\fh)$ should
  vanish at $o$ but here it clearly does not.} the definition
\eqref{eq:S@oexplicit} of $S$ to the whole of $\fg$ by $S_YX =
\nabla_XY$.  Together with invariance of the curvature, this allows
one to manipulate \eqref{eq:[F,A_0]}
\begin{equation*}
  \begin{split}
    [F,A_0]_{ab} &=  R( U_{\fm},(S_{U_{\fm}}+S_{U_{\fh}})Y_a,
    U_{\fm}, Y_b) + R(U_{\fm},Y_a, U_{\fm},(S_{U_{\fm}}+S_{U_{\fh}})Y_b ) \\
    &= \langle R(Y_b, U_{\fm}  ) U_{\fm},S_{U}Y_a\rangle + \langle
    R(U_{\fm},Y_a)U_{\fm} ,  S_{U}Y_b\rangle ~.
  \end{split}
\end{equation*}
Recall that $(A_0)_{ab} = -R(U_{\fm}, Y_a, U_{\fm}, Y_b)$, therefore,
we can take $F$ to be
\begin{equation*}
  F_{ab} = -\langle S_{U}(Y_a), Y_b\rangle = S(U, Y_b, Y_a)
\end{equation*}
where we have used that
\begin{equation*}
  \langle S_{U} Y_a , U_{\fm} \rangle =- c \langle
  Y_a, U_{\fm} \rangle = 0
\end{equation*}
and thus
\begin{equation*}
  \langle S_{U} Y_a , U_{\fm} \rangle \langle V ,
  R( U_{\fm},Y_a)U_{\fm}\rangle = \langle S_{U} Y_a , V \rangle
  \langle U_{\fm} , R( U_{\fm},Y_a)U_{\fm}\rangle = 0~.
\end{equation*}

In summary, the plane-wave limit is given by
\begin{equation*}
  \overline g = 2 e^{-2cx^+} dx^+ dx^-  + A_0\left(e^{-x^+F}\bx,
    e^{-x^+F}\bx\right)(dx^+)^2 + |d\bx|^2~,
\end{equation*}
where
\begin{equation}
  \label{eq:planewave_limit_formula}
  \begin{split} c &= -S(U,U,V) \\
    F_{ab} &= -S(U,Y_a,Y_b) \\
    (A_0)_{ab} &= -R(U_{\fm}, Y_a, U_{\fm}, Y_b)~,
  \end{split}
\end{equation}
with the curvature given by \eqref{eq:riemann} and the extension of
$S$ to $\fg$ given by
\begin{equation*}
  S(X,Y,Z) = \half \left< [X,Y]_\fm, Z_{\fm} \right> + \half \left<
    [Z,X]_\fm, Y_{\fm} \right> +\half \left< [Z,Y]_\fm, X_{\fm}
  \right>~.
\end{equation*}
The result is a non-singular plane-wave if $c=0$ and a singular
plane-wave if $c \neq 0$.

Notice that the often cumbersome enterprise of taking a plane-wave
limit is reduced, for the case of a homogeneous geodesic, to
straightforward algebraic calculations.

\subsection{The nearly-adapted method}

One thing the covariant approach to plane-wave limits teaches us, is
that the limit does not care about such details as the embedding of
the null geodesic \cite{CovariantPL}.  In particular, this means that
one should be able to use a not necessarily twist-free coordinate
system, which in many cases is the natural starting point, since
generically a geodesic vector will not be twist­free.

Let $\gamma$ be a null homogeneous geodesic generated by a
geodetic vector $U \in \fg$ so that equation \eqref{eq:geodetic}
holds. Let $V \in \fm$ be the dual null vector to $U_{\fm}$ and
complete with $(Y_i) \in \fm$ to a pseudo-orthonormal frame.

Let our local coset representative $\sigma$ be
\begin{equation*}
  \sigma = e^{\sum_i y^i Y_i} e^{vV} e^{uU} ~.
\end{equation*}
Then the Maurer--Cartan form $\theta$ can be expanded as
\begin{equation*}
  \sigma^*\theta = \theta^U U + \theta^V
  V + \theta^i Y_i + \theta^{\alpha} e_{\alpha}
\end{equation*}
where Greek indices are reserved for the isotropy and
$(e_{\alpha})$ is a basis for $\fh$. The metric can then be
expanded as
\begin{equation}
  \label{eq:metricMC}
  g = 2\theta^U \theta^V + \sum_i (\theta^i)^2 ~.
\end{equation}
Calculating the Maurer--Cartan form using $\sigma$ gives
\begin{equation*}
  \sigma^*\theta = \sigma^{-1} d \sigma =
  e^{-uU} e^{-vV} e^{-\sum_iy^i Y_i} d\left(e^{\sum_i y^i Y_i}
  \right) e^{vV} e^{uU} + e^{-uU} Vdv e^{uU} + Udu ~.
\end{equation*}
A few things are clear; first $du$ can only appear in $\theta^U$
and thus $\partial_u$ is null. This also tells us that the
isomorphism from the set of left invariant vector fields to the
Lie algebra $\fg$ that is determined by $\theta$ maps $\partial_u$
to $U$. We will denote the inverse of this isomorphism as $\fg \ni
X \mapsto X^*$ in the following.  Secondly,
\begin{equation*}
  \partial_u \theta^V = \partial_u \langle \theta_{\fm}
  , U_{\fm} \rangle = U^* g(\theta_{\fm}^* , U_{\fm}^*) =
  g(\nabla_{U^*} \theta_{\fm}^* , U_{\fm}^*)+ g(\theta_{\fm}^* ,
  \nabla_{U^*} U_{\fm}^*)~,
\end{equation*} where $\theta^*_{\fm} =\theta^U U^* + \theta^V
  V^* + \theta^i Y_i^*$.
Now applying the identity \eqref{eq:Koszul} we have,
\begin{equation*}
  \partial_u \theta^V= g([U^*, \theta_{\fm}^*], U_{\fm}^*) = -\langle
  [U,\theta_{\fm}]_{\fm}, U_{\fm}\rangle = -c \langle
  \theta_{\fm}, U_{\fm} \rangle = - c \theta^V ~,
\end{equation*}
where we have used that $U$ is geodetic. This shows that the only
dependence on $u$ in $\theta^V$ is a multiplicative factor of $e^{- c
  u}$. In particular, since the $dv$ part of $\theta$ is only
dependent on $u$, the $du dv$ part of the metric is of the form
$e^{-cu}$. This can be absorbed into the rest of the metric by a
coordinate change:
\begin{equation*}
  u \mapsto -\frac{1}{c} \log u ~,
\end{equation*}
however, this is not necessary since $u$ is not rescaled in the
plane-wave limit. Also, it is important to note that this coordinate
system is not necessarily a twist-free adapted coordinate system. We
will see that this is not important and one can still take a
plane-wave limit.

We can expand out the Maurer--Cartan form further and then take the
plane-wave limit.
\begin{equation*}
  \theta^U = du + \langle e^{-uU} V e^{uU}, V \rangle dv + \theta^U_i
  dy^i~,
\end{equation*}
where $\theta^U_i$ is a function of $u,v$ and $(y^i)$. Applying
the plane-wave limit rescaling $(u,v,y^i) \mapsto (u, \Omega^2 v,
\Omega y^i)$ to $\theta^U$ and taking the limit $\Omega \to 0$ we
see that $\theta^U \to du$.
\begin{align*}
  \theta^V &= e^{-cu}(dv +  \langle e^{-vV} e^{-\sum_iy^i Y_i}
  d\left(e^{\sum_i y^i Y_i} \right) e^{vV} , U_{\fm}\rangle dy^i) \\
  &=e^{-c u}(dv  + (\langle -y^j[Y_j,Y_i]_{\fm} + \cdots ,
  U_{\fm} \rangle dy^i)
\end{align*}
where $\cdots$ are terms involving $v$ and higher order terms in
$y^j$. If we rescale by $\Omega^{-2}$, apply the plane-wave limit
rescaling and take the limit $\Omega \to 0$ we find that all the
terms in $\cdots$ go to zero and we are left with,
\begin{equation*}
  \overline\theta^V = e^{-c u}(dv  -y^j\langle [Y_j,Y_i]_{\fm}, U_{\fm}
  \rangle dy^i) ~.
\end{equation*}
Similarly for $\theta^i$ we have
\begin{align*}
  \theta^i &= \langle e^{-vV} e^{-\sum_iy^i Y_i} d\left(e^{\sum_i y^i
      Y_i} \right) e^{vV} , Y_j\rangle dy^j \\
  &=\langle (e^{-uU} Y_j e^{uU})_{\fm} + \cdots , Y_i\rangle dy^j ~,
\end{align*}
where $\cdots$ are terms which involve $v$ and higher order terms
in $y^i$. Rescaling by $\Omega^{-1}$ and taking the plane-wave
limit we are left with
\begin{equation*}
  \overline\theta^i = \langle (e^{-uU} Y_j e^{uU})_{\fm}, Y_i\rangle dy^j
  ~.
\end{equation*}

Therefore the plane-wave limit of the metric in this coordinate
system is well defined:
\begin{equation*}
  \overline g = 2\overline\theta^Vdu +\sum_i (\overline\theta^i)^2 ~.
\end{equation*}
Expanding this we find that the metric is nearly a plane wave in Rosen
coordinates (as one would expect if this were the standard plane-wave
limit) but it has an additional $du dy^i$ term with a coefficient which
is linear in $y^j$:
\begin{equation*}
  \overline g = 2e^{-c u}du (dv  -y^j\langle [Y_j,Y_i]_{\fm}, U_{\fm}
  \rangle dy^i) + \langle (e^{-uU} Y_i e^{uU})_{\fm}, (e^{-uU} Y_j
  e^{uU})_{\fm}\rangle dy^idy^j ~.
\end{equation*}
Note that we have had to use that $U$ is geodetic in the
calculation of the last term.  We can make the change to a
Brinkmann type coordinate system irrespective of this extra term.
We find the metric in Brinkmann coordinates is
\begin{equation*}
  \begin{split}
    \overline g=& 2e^{-2cx^+}dx^- dx^+ - (\langle [U,Y_a]_{\fm}, Y_b \rangle-
    \langle [U,Y_b]_{\fm},Y_a \rangle - \langle [Y_a, Y_b]_{\fm},
    U_{\fm} \rangle) x^b dx^a dx^+  \\
    &+(\langle [U,Y_a]_{\fm}, [U,Y_b]_{\fm} \rangle - \langle
    [Y_a,[U,Y_b]]_{\fm},U \rangle)x^ax^b(dx^+)^2 + \sum_i (dx^i)^2 ~.
  \end{split}
\end{equation*}
Notice that
\begin{equation*}
  \langle [Y_a,[U,Y_b]]_{\fm},U \rangle
\end{equation*}
is symmetric in $a$ and $b$ because of the Jacobi identity and the
geodetic vector property \eqref{eq:geodetic}.  In light of the above,
we also define
\begin{equation}
  \label{eq:Fab}
  F_{ab} =  \frac{1}{2}\langle [Y_a, Y_b]_{\fm}, U_{\fm}
  \rangle - \frac{1}{2}\langle [U,Y_a]_{\fm}, Y_b \rangle +
  \frac{1}{2}\langle [U,Y_b]_{\fm},Y_a \rangle~.
\end{equation}

To show this is a plane wave and bring it to the proper Brinkmann
form we make the change of coordinates $\bx \mapsto e^{-x^+ F} \bx$,
which leaves the metric in the form
\begin{equation}
\label{eq:nearly_adapted_pl}
  2e^{-2cx^+}dx^- dx^+ + A_0 \left( e^{-x^+F}\bx, e^{-x^+F}\bx\right)
  (dx^+)^2 + |d\bx|^2 ~,
\end{equation}
where
\begin{equation}
  \label{eq:Aab}
  (A_0)_{ab}=\langle [U,Y_a]_{\fm}, [U,Y_b]_{\fm} \rangle - \langle
  [Y_a,[U,Y_b]]_{\fm},U \rangle + F^2_{ab} ~.
\end{equation}
An easy check shows that these formulae do indeed coincide with those
derived by the covariant method.

However, since we have not worked with an adapted coordinate
system, at no stage in the above discuss have we proved that the
formula we have obtained is actually applicable to the usual
plane-wave limit of the geodesic $\gamma$.  We will now remedy this
situation.

Consider a metric of the form
\begin{equation*}
  \label{gPlusLinTerm}
  2 du dv + \alpha dv^2 + \beta_i dy^i dv + K_{ij}y^i dy^j du +
  C_{ij}dy^idy^j ~,
\end{equation*}
such that $\partial_u$ is a null geodesic and $K_{ij}$ is
skew-symmetric.  Up to a coordinate transformation in $u$ this is
the form of the metric in equation \eqref{eq:metricMC}. An easy
calculation shows that the $R_{uiuj}$ component of the curvature
of this metric:
\begin{equation*}
  R(\partial_u,\partial_i)\partial_u = -\nabla_{\partial_u}
  \nabla_{\partial_i} \partial_u  + \nabla_{\partial_i}
  \nabla_{\partial_u} \partial_u +\nabla_{ [\partial_u,\partial_i]}
  \partial_u ~,
\end{equation*}
is independent of $K_{ij}$. If we apply the plane-wave limit
rescaling, multiply by $\Omega^{-2}$ and take the limit as $\Omega
\to 0$ we get
\begin{equation*}
  2 du dv +K_{ij}y^i dy^j du + C_{ij}(u)dy^idy^j ~.
\end{equation*}
This metric is a plane wave, as we can change to Brinkmann coordinates
and then absorb the linear term into the rest of the metric (as we did
above).  Since a plane wave is completely determined by the $R_{uiuj}$
part of its curvature, the metric \eqref{eq:nearly_adapted_pl} must be
isometric to the usual plane-wave limit of the geodesic $\partial_u$.

Let us consider in passing the plane-wave limits of the WZW model.
The algebraic data of a WZW model consists of a Lie algebra $\fg$
together with an nondegenerate invariant inner product.  The geometry
is therefore trivially that of a naturally reductive space which, as
we have seen above, means that every null vector in $\fg$ gives rise
to a homogeneous geodesic, hence every plane-wave limit preserves
homogeneity.  Moreover plane-wave limits are equivalent to contracting
$\fg$.  The above method for calculating the plane-wave limit is just
a manifestation of this fact: instead of expanding the Maurer--Cartan
forms in $\Omega$, one can redefine the generators $\widetilde U=U$,
$\widetilde V=\Omega^2V$ and $\widetilde E_a = \Omega E_a$ with a new
inner product $\langle \widetilde U, \widetilde V\rangle^\prime=1$ and
$\langle \widetilde E_a, \widetilde E_b\rangle^\prime =\eta_{ab}$.
This then means that we can write down a family of WZW models in terms
of $\langle -,-\rangle^\prime$ and the Lie bracket---equivalently, the
homogeneous structure---that interpolates between the original model
at $\Omega =1$ and its plane-wave limit at $\Omega =0$.  It is then
clear that the Poisson bracket for this family has a regular limit,
and that group contraction is extended to a contraction of the
associated affine algebra and also of the Yangian \cite{Abdalla}.
This means that the diagram of contraction and quantisation in
\cite{Penati} should indeed commute.

\subsection{Homogeneous structures under the plane-wave limit}
\label{sec:homstructpl}

There are circumstances where one can say more than that the
plane-wave limit is homogeneous, and in addition say something about
the type of homogeneous structure inherited by the plane-wave limit
along a homogeneous geodesic. It is clear from
\eqref{eq:planewave_limit_formula} that the homogeneous structure of
the plane-wave limit along a homogeneous geodesic is inherited from
the original metric $g$ in some sense, since the whole plane-wave
limit metric is defined in terms of the algebraic data. In fact,
\eqref{eq:planewave_limit_formula} for $F$ can be interpreted as the
Ambrose--Singer formula $\overline\nabla \, \overline R = \overline S
\cdot \overline R$ on the plane-wave limit.  However, this homogeneous
structure is not inherited continuously in the limit, so it is
difficult to draw conclusions about the type of homogeneous structure
inherited under the plane-wave limit.  To study this situation we may
consider a stronger form of heritability of the homogeneous structure,
namely when this is inherited continuously in the limit.

On a reductive homogeneous space, the Ambrose--Singer theorem provides
us with a connection, namely $\widetilde{\nabla}$, relative to which
the metric $g$, the Riemann tensor $R$ and the homogeneous structure
$S$ are parallel.  One way to guarantee the heritability of
homogeneity would be to have a well-defined limit of
$\widetilde{\nabla}$ or rather, since $\nabla$ has a well-defined
limit, to have a well-defined limit of the homogeneous structure.

When $\widetilde{\nabla}$ has a well-defined limit, then also its
curvature has a well-defined limit and, seeing the discussion around
equation \eqref{eq:Salgebra}, one must conclude that the plane-wave
limit is equivalent to a Inönü--Wigner contraction, where the extra
isometries that can arise through the plane-wave limit will be
elements of the isotropy subalgebra.  A result mentioned in
\cite{SimonPL} gives a sufficient and necessary criterion for this to
happen: there exists a well-defined plane-wave limit of $S$ if and
only if the geodesic along which the limit is performed is canonically
homogeneous. Since on a symmetric space all geodesics are canonically
homogeneous, this provides an \emph{a posteriori} explanation for the
results in \cite{HatKamiSaka}.  Furthermore, on the Kaigorodov space
there are two canonically homogeneous geodesics, and as such the
identification in \cite[Section~3.3]{Patricot}, albeit through a
different contraction, can be reproduced. One can easily check that
this leads to the statement that the resulting singular plane wave is
also a group manifold with a left-invariant metric.

Let us then have a better look at the plane-wave limit of $S$.
Since the torsion in the plane-wave limit must scale as the
Levi-Cività connection, it is evident that the plane-wave limit of
$S$ is given by
\begin{equation*}
  \overline S = \lim_{\Omega\rightarrow 0} \Omega^{-2}
  \half S(\tilde{x})_{ABC} d\tilde y^A\otimes d\tilde y^B \wedge
  d\tilde y^C~,
\end{equation*}
where $\tilde y=\left(u,\Omega^2 v,\Omega y^i\right)$.  It then
follows that the condition for the existence of a non-singular limit
is the regularity of $\lim_{\Omega\rightarrow 0}
\Omega^{-1}S_{uui}(\tilde{y})$, which together with $S_{uuv}$,
$S_{uij}$ and $S_{iuj}$, are the components surviving in the
plane-wave limit.

We are now in a position to see why the regularity of the
plane-wave limit implies that the geodesic must be canonically
homogeneous.  Indeed, if we let $U$ be the geodetic vector, so that
$\nabla_{U}U=0$, then
\begin{equation*}
  S_{uui} = S(U,U,\partial_i) = -g ( \partial_i,
  \widetilde\nabla_UU)~.
\end{equation*}
Decomposing $\widetilde\nabla_UU = A U + B^i\partial_i$, where a
$\partial_v$ contribution is impossible due to the fact that
$S(U,U,U)=0$, one sees that $S_{uui}= - C_{ij}B^j$. This implies that
$B^j=0$ if and only if a regular plane-wave limit of $S$ exists. At
the same time this implies $\widetilde\nabla_UU \propto U$, or rather
the geodesic is also a geodesic of the canonical connection, whence
the geodesic is canonically homogeneous.

A first thing to observe is that the component $S_{uui}$ cannot be
part of $\eT_1$ nor of $\eT_3$, from which we can conclude that the
plane-wave limit of a $\eT_1\oplus \eT_3$ structure is regular. This
reinforces the discussion about the existence of canonically
homogeneous geodesics on spaces admitting a $\eT_1\oplus \eT_3$
structure in Section~\ref{sec:planewavelimits}.

We can be a bit more precise as to what homogeneous structure the
plane-wave limit of a $\eT_1 \oplus \eT_3$ space will inherit, by
applying the formula \eqref{eq:planewave_limit_formula}.  Consider a
null geodesic $\gamma$ generated by the Killing vector $U \in \fm$ and
dual null vector $V \in \fm$. Then, introducing $\alpha(Z) = g(\xi,Z)$
as in equation \eqref{eq:planewave_limit_formula},
\begin{equation*}
  c= -S(U,U,V) = \alpha(U)~.
\end{equation*}
There are two scenarios to consider, i) $\alpha(U) = 0$ and ii)
$\alpha(U) \neq 0$.  The first case corresponds to the original space
$g$ being naturally reductive and the second case to $g$ being a
singular homogeneous plane-wave.  Comparing this with the
classification of homogeneous plane-waves reviewed in
Section~\ref{sec:HSwaves}, we must conclude that in case i) the
resulting spacetime admits a pure $\eT_{3}$ structure and must be a
regular homogeneous plane wave, whereas in case ii) the resulting
spacetime is a singular homogeneous plane wave.  Note that the
plane-wave limit of a plane-wave is not necessarily trivial, only if
the limit is along the defining null geodesic of the wave.

In general not much more can be said about which classes of reductive
homogeneous spaces guarantee the existence of canonically homogeneous
null geodesics. This is exemplified by the next three examples, all of
which are discussed in more detail in Section \ref{sec:examples}. The
first example is of course the Kaigorodov space, for which there
exists a unique homogeneous structure of generic type, for example,
equation \eqref{eq:HSkaigorodov}. Evaluating this homogeneous
structure in the plane-wave limit one can see that it is well defined
if and only if the limit is taken along a null geodesic with initial
direction given by \eqref{eq:kaigorodov_nullvecs} with $\alpha^2=1$.
These are just the cases that characterise the null homogeneous
geodesics and lead to a singular homogeneous plane wave or flat space.

The above example shows that if a spacetime has a unique
homogeneous structure containing a $\eT_{2}$ contribution, then
generically there are geodesics such that homogeneity is lost in
the plane-wave limit.  The next example shows that this also holds
for cases where we have a family of homogeneous structures.
Consider the Komrakov 1.4.6 metric \cite{Komrakov}
\begin{equation*}
  e^{-2y} \left( 2du\left[ dv + ydu\right] + dx^2 \right) +
  dy^2~.
\end{equation*}
As shown in Section~\ref{sec:komrakov146}, this homogeneous metric
admits a two-parameter family of generic homogeneous structures, which
minimally is of type $\eT_1\oplus \eT_2$.  The only plane-wave limit
for which the resulting spacetime is homogeneous is in the $v$
direction, which incidentally also corresponds to the only
(canonically) homogeneous null geodesic, leading to flat
space.

Seeing the above examples, one might be tempted to conclude that if
the homogeneous structure contains a $\eT_2$ contribution, then there
are geodesics such that the plane-wave limit is non-homogeneous.  In
order to show that this is certainly not the case, consider the metric
(see Section~\ref{sec:kaplan})
\begin{equation}
  \label{eq:KaplanMet}
  ds^{2}  = -\left( dt - x^1dx^4 - x^2dx^3\right)^2
  + \left( dy + x^2dx^4 - x^1 dx^3 \right)^2
  + dx_idx^i~,
\end{equation}
which is a lorentzian version of Kaplan's first example of a
g.o.~space, that is in no way naturally reductive (see, for example,
\cite{KowNik,KowNikApp}).  A small calculation of the homogeneous
structure shows that it admits a 3-parameter family of
$\eT_2\oplus\eT_3$ structures, which does not contain a pure $\eT_3$
point.  Since this space is g.o., every (null) geodesic is homogeneous
and we are guaranteed that the plane-wave limit is homogeneous. There
are however only two canonically homogeneous null geodesics, namely
those along $X_5 \pm X_6$, and for those the resulting spacetime is
flat space and the limit of $S$ is regular.  For the remaining null
geodesics, the result of the plane-wave limit is a homogeneous plane
wave, but the plane-wave limit of $S$ is not well-behaved; this is due
to the fact that although the null geodesics are homogeneous, there is
no reductive split such that a given null geodesic becomes canonical.

The conclusion of the examples then must be that on a reductive
homogeneous space whose homogeneous structure always contains a
$\eT_{2}$ contribution, there exist plane-wave limits along which the
homogeneous structure is singular.  This, however, need not imply the
absence of homogeneous geodesics and thus loss of homogeneity in the
plane-wave limit, as evidenced by the example of Kaplan's g.o.~space.

Given a reductive homogeneous space $(M,g)$ which admits a
homogeneous structure of a type other than $\eT_{1} \oplus \eT_3$
and given a canonically homogeneous geodesic $\gamma$ of $(M,g)$,
what can one say about the homogeneous plane-wave limit along
$\gamma$? Let us consider the other two types of homogeneous
structures, $\eT_{1} \oplus \eT_2$ and $\eT_{2} \oplus \eT_3$,
separately. First, suppose that $g$ admits a homogeneous structure
$S$ of type $\eT_{2} \oplus \eT_3$, that is $S$ is in the kernel
of the map $C$ defined in \eqref{eq:cmap}. Then the plane-wave
rescaling $S_{\Omega}$ will be in the kernel of $C_{\Omega}$, and
hence a homogeneous structure of type $\eT_{2} \oplus \eT_3$ for
$g_{\Omega}$ (as defined in Section~\ref{sec:planewavelimits}). If
we can define a family of pseudo orthonormal frames
$(e_a(\Omega))$ for $g_{\Omega}$, such that in the limit $\Omega
\to 0$ the frame $(e_a(0))$ is a well-defined pseudo orthonormal
frame for $\overline g$, then continuity will ensure that $\overline S$ is
in the kernel of $\overline C$.

We can exhibit such a basis by applying the Gram--Schmidt process to
the frame $(\partial_{y^1}, \dots, \partial_{y^{n-2}}, \partial_u -
\partial_v, \partial_u + \partial_v)$ for \eqref{eq:twistfree}, starting
from the left. This gives an orthonormal frame $(e_1(\Omega), \dots,
e_{n-2}(\Omega))$ for the transverse $y^i$ coordinates, together with
\begin{align*}
  e_u(\Omega) &= \frac{1}{(2-\Omega^2 \alpha- \sum_{i=1}^{n-1}
    \Omega^2 \eta_i^2)} \left(  \frac{\partial}{\partial u} -
    \frac{\partial}{\partial v} + \Omega \sum_{i=1}^{n-1} \eta_i e^i
  \right), \\
  e_v(\Omega) &= \frac{1}{(2+ \Phi(\Omega))} \left(
    \frac{\partial}{\partial u} + \frac{\partial}{\partial v} +
    \Omega^2 \alpha e_u(\Omega)- \Omega \sum_{i=1}^{n-1} (\eta_i e^i
    + \Omega^2 \eta^2_i e^u) \right) ~,
\end{align*}
where $\eta_i := g(\frac{\partial }{\partial v}, e^i)$ and
$\Phi(\Omega)$ is some function of $(\Omega, u,v,y^k)$ such that
$|e^v|^2 = 1$ and which tends to zero as $\Omega \to 0$.  Taking the
limit $\Omega \to 0$, we obtain an orthonormal basis with respect to
$\overline g$, namely $( e_1, \dots , e_{n-2},( \partial_u - \partial_v
)/2, (\partial_u + \partial_v )/2 )$.

Consequently, we find that $\overline S$ is of type $\eT_{2} \oplus
\eT_3$. Since homogeneous plane-waves are essentially of type $\eT_{1}
\oplus \eT_3$, this gives a further restriction on $\overline S$:
\begin{equation*}
  0  =C(\overline S)(\partial_v)=\overline S(e_u,e_u,\partial_v)
  +\overline S(e_v,e_v,\partial_v)+\overline S(e_i,e_i,\partial_v) =
  \frac{1}{2} \overline S_{uuv}~,
\end{equation*}
where we have used $\overline S_{vv}^u = \overline S_{e_ie_i}^u = 0$. We have
already seen \eqref{eq:sing_hom_wave_hom_struct} that $\overline S_{uuv}$
is non-zero for the singular plane-waves. Therefore, the plane-wave
limit must be a non-singular plane-wave.

If $g$ admits a homogeneous structure of type $\eT_{1} \oplus \eT_2$,
so that $S^{\,\yng(1,1,1)}$ vanishes, then one can merely say that the
plane-wave limit also admits a homogeneous structure of this type (by
continuity).  This homogeneous structure $\overline S$ may not be of type
$\eT_{1} \oplus \eT_3$ since the plane-wave may admit many different
homogeneous structures.  In addition, the formulae
\eqref{eq:planewave_limit_formula} shed little light on the subject.

\section{Examples}
\label{sec:examples}

In this section we discuss several homogeneous spacetimes in detail
and discuss their homogeneous structures. We also compute their
plane-wave limits using our Lie algebraic formulation.  First we
summarise the methodology by which we explore the possible plane-wave
limits.

\subsection{Methodology}
\label{sec:method}

Given a homogeneous space in terms of the following data: a
Lie algebra of isometries $\fg = \fh \oplus \fm$ with an
$\fh$-invariant lorentzian inner product $\left<-,-\right>$ on $\fm$,
we follow the following procedure:
\begin{enumerate}
\item We first determine the orbit decomposition of the projectivised
  light-cone of $\fm$ under the exponentiated action of $\fh$. This
  will determine the possible null directions up to isometry.  In
  practise we label these orbits by giving a null direction in each
  orbit.
\item For each such null direction $u \in \fm$ we determine whether
  the null geodesic pointing along $u$ is homogeneous.  In other
  words, we determine whether there is some $X \in \fh$ for which
  $U:=u + X$ is geodetic; that is, whether $U$ obeys
  \eqref{eq:geodetic} for some value of $c$.  If it does, then the
  plane-wave limit along $U$ will be homogeneous: regular if $c=0$ and
  singular otherwise.
\item Finally we determine the explicit form of the plane-wave
  metric.  To do this we choose a frame $u,V,Y_a$ for $\fm$ such that
  $\left<u,V\right> = 1$ and $\left<Y_a,Y_b\right> = \delta_{ab}$ and
  then compute the matrices $F$ and $A$ using formulae~\eqref{eq:Fab}
  and \eqref{eq:Aab}, respectively.
\end{enumerate}
Clearly many of these steps can be performed (or at least checked)
using one's favourite computer algebra software.

\subsection{Kaigorodov spaces}
\label{sec:kaigorodov}

The Kaigorodov space $K$ is an ($n+3$)-dimensional lorentzian manifold
with metric \cite{CLPimf}
\begin{equation*}
  -(\theta^0)^2 + \sum_{i=1}^{n+2} (\theta^i)^2
\end{equation*}
where
\begin{equation*}
  \label{eq:kaigodorov_metric}
  \theta^0 = e^{(4+n)\ell\rho} dt~, \quad
  \theta^i = e^{2\ell \rho} dy^i~, \quad
  \theta^{n+1} = e ^{-n \ell\rho} dx + e^{(4+n)\ell\rho} dt~, \quad
  \theta^{n+2} = d\rho~,
\end{equation*}
where, here and in the sequel, the indices $i,j,...$ run from $1$ to
$n$.  This spacetime can be seen to have a pp-wave singularity and is
not geodesically complete \cite{Podolsky}.  Up to homothety, we can
(and will) set $\ell =1$ from now on.

We observe that the $\theta^a$ form a differential ideal with
structure \emph{constants}:
\begin{align*}
  d\theta^0 &= (4+n) \theta^{n+2} \wedge \theta^0\\
  d\theta^i &= 2  \theta^{n+2} \wedge \theta^i\\
  d\theta^{n+1} &= - n  \theta^{n+2} \wedge \theta^{n+1} + 2
  (2+n)  \theta^{n+2} \wedge \theta^0\\
  d\theta^{n+2} &= 0~.
\end{align*}
This means that the dual vector fields $X_a$, defined by
$\theta^a(X_b) = \delta^a_b$, form a Lie algebra, denoted $\fk$.
We can read off their Lie brackets from the above differentials
\begin{align*}
  [X_{n+2}, X_0] &= - (4+n) X_0 - 2 (2+n)  X_{n+1}\\
  [X_{n+2}, X_{n+1}] &= n X_{n+1}\\
  [X_{n+2}, X_i] &= - 2 X_i~.
\end{align*}
Notice that from the expression of the metric, these vector fields
form a pseudo-orthonormal frame.  It is convenient to diagonalise the
adjoint action of $X_{n+2}$ by redefining $X_0 \mapsto X_0 +
X_{n+1}$.  We now have the simpler brackets
\begin{align*}
  [X_{n+2}, X_0] &= - (4+n) X_0\\
  [X_{n+2}, X_{n+1}] &= n X_{n+1}\\
  [X_{n+2}, X_i] &= - 2 X_i~,
\end{align*}
at the price that the metric in this new basis is no longer diagonal,
but instead $X_0$ is now null and $\left< X_0, X_{n+1} \right> = 1$.

In summary, we have exhibited the Kaigorodov space $K$ as a Lie group
with a left-invariant lorentzian metric.  In particular it is
trivially reductive.  The corresponding homogeneous structure $S_{abc}
= S(X_a,X_b,X_c)$, from equation \eqref{eq:S@oexplicit}, is given by
\begin{equation}
  \label{eq:HSkaigorodov}
  \begin{aligned}[m]
    S_{n+2,0,n+1} &= - (2+n)\\
    S_{n+1,n+1,n+2} &= n
  \end{aligned}
  \qquad\qquad
  \begin{aligned}[m]
    S_{n+1,0,n+2} &= S_{0,n+1,n+2} = -2\\
  S_{i,j,n+2} &= -  \delta_{ij}~.
  \end{aligned}
\end{equation}
It is not hard to see that it has generic type $\eT_1 \oplus \eT_2
\oplus \eT_3$.

The full isometry Lie algebra of $K$ is larger than the Lie algebra
$\fk$ generated by the $X_a$.  Indeed, it has in addition an
$\fiso(n)$ Lie algebra, with generators $L_{ij}$ and $L_i$, with
the $\fso(n)$ generators $L_{ij}$ acting on the $X_i$ as vectors and
together with the following brackets
\begin{align*}
  [L_i,X_j] &= - \delta_{ij} X_0 \\
  [L_i,X_{n+1}] &= X_i \\
  [X_{n+2}, L_i] &= -(n+2) L_i~.
\end{align*}
It is clear from this last bracket that this is not a reductive split;
although we still have an action of $\fiso(n)$ on the tangent space
$T_oK$ at the identity by projecting the Lie bracket to $\fk$.

We now determine the action of the isotropy group $\ISO(n)$ on the
celestial sphere in $T_oK$.  Relative to the ordered basis
$(X_1,\ldots,X_{n+2},X_0)$, the typical element $(A,\bbee)$ of
$\ISO(n) = \SO(n) \ltimes \RR^n$ has matrix
\begin{equation*}
  \begin{pmatrix}
    A & A\bbee & \bzero & \bzero \\
    \bzero & 1 & 0 & 0 \\
    \bzero & 0 & 1 & 0 \\
    -\bbee^t & -\half |\bbee|^2 & 0 & 1
  \end{pmatrix}
\end{equation*}
which has been obtained as the product
\begin{equation*}
  \begin{pmatrix}
    A & \\
    & \1_3
  \end{pmatrix} \exp(b^i L_i)~.
\end{equation*}
Acting on a tangent vector $v = (\bv, v^{n+1}, v^{n+2}, v^0) \in
T_oK$, we find
\begin{equation*}
  (A,\bbee) \cdot
  \begin{pmatrix}
    \bv\\ v^{n+1}\\ v^{n+2}\\ v^0
  \end{pmatrix}
  =
  \begin{pmatrix}
    A \bv + v^{n+1} A \bbee\\
    v^{n+1}\\
    v^{n+2}\\
    v^0 - \bbee^t \bv - \half |\bbee|^2 v^{n+1}
  \end{pmatrix}~.
\end{equation*}
Let $v$ have zero norm, so that
\begin{equation*}
  (v^{n+1})^2 +   (v^{n+2})^2 +  |\bv|^2 = - 2 v^0 v^{n+1}~.
\end{equation*}
Since $v\neq 0$, it follows that $v^0 \neq 0$.  We must therefore
distinguish two cases, according to whether $v^{n+1}$ does or does not
vanish.
\begin{itemize}
\item If $v^{n+1} =0$, then also $v^{n+2}=0$ and $\bv = \bzero$.  We
  can then choose $v^0 = 1$, whence $v = X_0$.
\item If $v^{n+1} \neq 0$, then we can choose $\bbee = - \bv/v^{n+1}$
  to bring $v$ to the form
  \begin{equation*}
    \begin{pmatrix}
      \bzero \\ v^{n+1} \\ v^{n+2} \\ -\frac{1}{2v^{n+1}} ((v^{n+1})^2
      + (v^{n+2})^2)
    \end{pmatrix}~,
  \end{equation*}
  where we have used that $v$ is null.  We can choose $v^{n+1} = 1$,
  $v^{n+2} = \alpha$ so that finally
  \begin{equation*}
    v = X_{n+1} + \alpha X_{n+2} - \half (1+\alpha^2) X_0~.
  \end{equation*}
\end{itemize}

In summary, we have two possible null directions up to the action of
the isotropy subgroup, one of them parametrised by a real number
$\alpha$:
\begin{equation}
  \label{eq:kaigorodov_nullvecs}
  X_0 \qquad\text{and}\qquad X_{n+1} + \alpha X_{n+2} - \half
  (1+\alpha^2) X_0~.
\end{equation}

One checks that $X_0$ is a geodetic vector with $c=0$, and that
the null geodesic along $X_{n+1} + \alpha X_{n+2} - \half (1+\alpha^2)
X_0$ is homogeneous only when $\alpha^2 = 1$, in which case $X_{n+1} +
\alpha X_{n+2} - X_0$ is geodetic with $c=-\alpha(4+n)$.  In the first
case, therefore, the corresponding plane-wave limits will be a regular
homogeneous plane wave, whereas in the second case the limit will be a
singular homogeneous plane wave.

It is not difficult to see that in both cases the skew-symmetric
matrix $F$ given by equation~\eqref{eq:Fab} vanishes.  This means the
limit is a symmetric plane wave, whose metric is determined by the
symmetric matrix $A$ in equation~\eqref{eq:Aab}.  It is easy to show
that in the first case, where the geodetic vector is $X_0$, the
symmetric matrix $A=0$, whence the plane-wave limit is flat.  In the
case where the geodetic vector is $X_{n+1} + \alpha X_{n+2} - X_0$, a
calculation shows that the nonzero components of $A$ are
\begin{equation*}
  A_{ij} = 4 \delta_{ij} \qquad\text{and}\qquad A_{n+1,n+1} = n^2~.
\end{equation*}

\subsection{Higher-dimensional Gödel universes}
\label{sec:HSGoedel}

The five-dimensional Gödel universe is a reductive spacetime and also
a maximally supersymmetric solution of minimal five-dimensional
supergravity, whose M-theory lift preserves 20 supersymmetries
\cite{GGHPR}.  The plane-wave limit of the five-dimensional Gödel
universe is the five-dimensional maximally supersymmetric plane wave
\cite{Meessen}.  The plane-wave limits of the M-theory Gödel universe
were investigated in \cite{BMO} and shown to give rise to a family of
time-dependent plane waves interpolating between two Cahen--Wallach
spaces, one of which corresponds to the M-theory lift of the
five-dimensional maximally supersymmetric plane wave.  In this
subsection, we will rederive these results using our Lie algebraic
formalism.

\subsubsection{The five-dimensional Gödel universe}

We start with the five-dimensional Gödel universe, which is defined on
a circle bundle over flat euclidean space:
\begin{equation}
  \label{eq:metric}
  g = - (dt + A)^2 + \sum_{i=1}^4 (dx^i)^2~,
\end{equation}
where the connection one-form $A$ is given by
\begin{equation}
  \label{eq:oneform}
  A = \half (x^1 dx^2 - x^2 dx^1) - \half (x^3 dx^4 - x^4 dx^3)~.
\end{equation}
The two-form $F$ is given simply by
\begin{equation}
  \label{eq:twoform}
  F = dA = dx^1 \wedge dx^2 - dx^3 \wedge dx^4~,
\end{equation}
which is clearly an anti-selfdual two-form in $\EE^4$ with respect to
the natural orientation.  Any infinitesimal symmetry of $F$ can be
promoted to an isometry by adding a compensating gauge transformation.
The two-form $F$ is manifestly invariant under a subgroup $\U(2)
\ltimes \RR^4$ of the euclidean group of isometries of $\EE^4$ and in
addition by the $\U(1)$ group of translations along the fibre
generated infinitesimally by $\d_t$.  Clearly $\U(2)$ and the fibre
$\U(1)$ are still invariances of the metric, but the translations are
not because they do not leave $A$ invariant.  Nevertheless they can be
corrected to make $dt + A$ and hence the metric invariant.  Doing so
one finds the following Killing vectors leaving $g$ and $F$ invariant:
\begin{equation}
  \label{eq:isometries}
  \begin{gathered}
    \d_t \qquad \d_1 - \half x^2 \d_t \qquad \d_2 + \half x^1 \d_t
    \qquad
    \d_3 + \half x^4 \d_t \qquad \d_4 - \half x^3 \d_t\\
    x^1 \d_2 - x^2 \d_1 \qquad x^3 \d_4 - x^4 \d_3\\
    x^1 \d_3 - x^3 \d_1 + x^2 \d_4 - x^4 \d_2 \qquad x^1 \d_4 - x^4
    \d_1 - x^2 \d_3 + x^3 \d_2~.
  \end{gathered}
\end{equation}
Notice that at any point $(t,x^i)$ of $M$, the five Killing vectors in the
first line span the tangent space, so that $M$ is indeed a homogeneous
space.

The isometry algebra $\fg$ is isomorphic to the semidirect product
\begin{equation*}
  \fg \cong \left( \fsu(2) \times \fu(1) \right) \ltimes \fh(4)~,
\end{equation*}
where $\fh(4)$ is the five-dimensional Heisenberg algebra
\begin{equation*}
  [P_i, P_j] = \Omega_{ij} P_0~,
\end{equation*}
generated by $P_0 = \d_t$ and $P_i = \d_i - \half \sum_j \Omega_{ij}
x^j \d_t$, where $\Omega_{ij}$ is the symplectic form with nonzero
entries $\Omega_{12} = 1 = - \Omega_{21}$ and $\Omega_{34} = -1 = -
\Omega_{43}$.  In the above expression for $\fg$, $\fsu(2) \times
\fu(1) \subset \fso(4)$ acts on $\fh(4)$ by restricting the natural
action of $\fso(4)$ under which $P_0$ is a scalar and $P_i$ is a
vector.  The corresponding isometry group $G$ is given by
\begin{equation*}
  G \cong \U(2) \ltimes \H(4)~,
\end{equation*}
with $\U(2) \subset \SO(4)$ acting on $\H(4)$ in the natural way.

Let $o \in M$ be the point with coordinates $(t=x^i=0)$.   The
vectors $P_0,P_1,\ldots,P_4$ form a pseudo-orthonormal frame for
$T_oM$, with $P_0$ timelike.  The little group of $o$ is precisely the
natural $\U(2)$ subgroup of $G$.  Therefore $M = G/\U(2)$ is the
$G$-orbit of $o$.  From the classification of lorentzian symmetric
spaces in \cite{CahenWallach} or from a direct calculation, it follows
that $M$ is not symmetric.

We can also see this by exhibiting the homogeneous structures of the
Gödel universe.  Considering the reductive split $\fg = \fh \oplus
\fm$ with $\fg$ the full isometry algebra, we find using equation
\eqref{eq:S@oexplicit} that the components $S_{abc} = S(P_a,P_b,P_c)$
of the homogeneous structure at $o$ are given by
\begin{equation*}
  S_{0ij} = S_{i0j} = - S_{ij0} = \half \Omega_{ij}~,
\end{equation*}
which can be seen to be of type $\eT_2 \oplus \eT_3$.

We can deform this homogeneous structure by considering a reductive
split $\fg = \fh \oplus \fm'$ where $\fm'$ is the graph of an
$\fh$-equivariant linear map $\fm \to \fh$.  Decomposing $\fm$ and
$\fh$ into irreducibles we find that there is a one-parameter map of
such linear maps $\varphi_\alpha (v^i P_i) = \alpha v^0 Y_0$, where
$Y_0 \in \fh$ is the Killing vector $Y_0 = x^1 \d_2 - x^2 \d_1 + x^3
\d_4 - x^4 \d_3$.  Its graph $\fm'$ is spanned by
\begin{equation*}
  P_1~, \quad P_2~, \quad P_3~, \quad P_4~, \quad \text{and} \quad
  P_0 + \alpha Y_0~.
\end{equation*}
This modifies the $[-,-]_{\fm'}$ brackets:
\begin{equation*}
  [P_i,P_j]_{\fm'} = \Omega_{ij} (P_0 + \alpha Y_0)
  \qquad\text{and}\qquad
  [P_0+ \alpha Y_0, P_i]_{\fm'} = \alpha \Omega_{ij} P_j~.
\end{equation*}
We can now compute the corresponding homogeneous structure using
formula \eqref{eq:S@oexplicit} and we obtain a one-parameter family of
$\eT_2 \oplus \eT_3$ structures:
\begin{equation*}
  S_{0ij} = (\half + \alpha) \Omega_{ij}\qquad\text{and}\qquad S_{i0j}
  = - S_{ij0} = \half \Omega_{ij}~.
\end{equation*}
Naturally, when $\alpha = 0$ we recover the earlier homogeneous
structure.  Clearly for generic $\alpha$ we have a homogeneous
structure of type $\eT_2 \oplus \eT_3$, but for $\alpha = -1$ it is of
type $\eT_3$ and for $\alpha = \half$ it is of type $\eT_2$.

One can obtain more homogeneous structures by considering smaller
subalgebras, but we will not do so here.

In order to determine the possible plane-wave limits of the Gödel
universe we will exploit the covariance property of the plane-wave
limit \cite{Limits}.  This says that if two null geodesics in $M$ are
related by an isometry of $M$, then the corresponding plane-wave
limits are themselves isometric.  A null geodesic $\gamma$ in $M$ is
locally determined by the following data: an initial point $\gamma(0)
\in M$ and an initial direction $\gamma'(0)$, which is a point on the
future-pointing, say, celestial sphere at $\gamma(0)$.  Since $M$ is
homogeneous, we can choose $\gamma(0)$ to be any convenient point.  We
will choose the point $o$ above with coordinates $(t=0,x^i=0)$ and
retain the freedom of using the isotropy subgroup of $o$.  The
(future) celestial sphere at $o$ consists of those vectors $v = v^\mu
P_\mu$ such that $\left<v,v\right> = 0$ and $v^0 = 1$, which is the
unit three-sphere in $\EE^4 = \left<P_0\right>^\perp$.  The isotropy
group $\U(2)$ acts on $\EE^4$ by restricting the natural
representation of $\SO(4)$, whence it acts transitively on the
spheres.  Therefore we see that the isometry group of $(M,g,F)$ acts
transitively on the space of null geodesics and hence all plane-wave
limits are isometric.

Let us choose our geodesic to point in the direction of $P_0 + P_1$.
This vector is not geodetic, since it does not satisfy equation
\eqref{eq:geodetic} for any value of $c$.  We modify it by adding a
vector $X \in \fh$ in such a way that \eqref{eq:geodetic} is
satisfied.  A quick calculation shows that $X = -Y_0$ does the trick.
The resulting Killing vector $P_0 + P_1 - Y_0$ is geodetic with $c=0$.
This means that the plane-wave limit will be a regular homogeneous
wave.  Moreover, we see that this is a canonically homogeneous
geodesic, since as shown above there is a reductive split $\fg = \fh
\oplus \fm'$ with $\fm'$ spanned by $P_0 - Y_0$, $P_1$, $P_2$, $P_3$
and $P_4$.

In fact, as we now show, the limit is a symmetric plane wave, with
geometry a five-dimensional Cahen--Wallach space.  This vacuum of
minimal five-dimensional supergravity was discovered in
\cite{Meessen}.  To determine the limit we employ the formulae
\eqref{eq:Fab} and \eqref{eq:Aab}.  We find that
\begin{equation*}
  F =
  \begin{pmatrix}
    0 & 0 & 0 \\ 0 & 0 & -\half \\ 0 & \half & 0
  \end{pmatrix}
  \qquad\text{and}\qquad
  A_0 =
  \begin{pmatrix}
    -1 & 0 & 0 \\ 0 & -\frac14 & 0 \\ 0 & 0 & -\frac14
  \end{pmatrix}~,
\end{equation*}
in agreement with the results of \cite{BMO}.

\subsubsection{The Gödel universe in M-theory}

The five-dimensional Gödel universe can be lifted to a supersymmetric
M-theory background $(\widetilde M,g,G)$ preserving 20 supersymmetries
\cite{GGHPR} simply by taking its cartesian product with a flat
six-dimensional space.  It is convenient to think of this
six-dimensional space as $\CC^3$ with its standard Kähler structure
$\omega$, whence $\widetilde M = M \times \CC^3$ metrically.  The
M-theory four-form is then $G = F \wedge \omega$, whence the symmetry
group of this M-theory background is
\begin{equation*}
  \left(\U(2) \ltimes \H(4)\right) \times \left( \U(3) \ltimes \RR^6
  \right)~,
\end{equation*}
which still acts transitively, making $(\widetilde M,g,G)$ into a
homogeneous background.  Let $z^\alpha$ denote local coordinates on
$\CC^3$ and let $o$ be the point on the eleven-dimensional product
manifold with coordinates $t=x^i = z^\alpha = 0$.  The isotropy
subgroup at this point is $\U(2) \times \U(3)$, which is reductive.

The isotropy subgroup acts with cohomogeneity one in the (future)
celestial sphere in $T_o\widetilde M$.  Indeed, any tangent vector
decomposes into $\bv = \bv_G + \bv'$, with $\bv_G$ the component
tangent to the five-dimensional Gödel universe and $\bv'$ the
component tangent to $\CC^3$.  The isotropy subgroup preserves the
norms $|\bv_G|^2$ and $|\bv'|^2$ separately.  Let $\bv$ be a
future-pointing null vector.  By further rescaling, we can ensure that
its $P_0$ component is $1$, whence $\bv_G = P_0 + \bv_\perp$ where
$|\bv_\perp|^2 + |\bv'|^2 = 1$.  Fix an angle $\vartheta \in
[0,\frac{\pi}2]$ and let $|\bv_\perp| = \cos \vartheta$ and $|\bv'| =
\sin\vartheta$.  The isotropy subgroup cannot change $\vartheta$, but
it acts transitively on these spheres, whence we can make $\bv_\perp$
and $\bv'$ point in any desired direction.  Letting $T_i$ denote the
translation generators for the $\RR^6$ subgroup of the isometries of
$\CC^3$, we can write the null vector as
\begin{equation*}
  P_0 + \cos\vartheta P_1 + \sin\vartheta T_1~.
\end{equation*}
This vector is not geodetic, however, unless we add $-Y_0$, as in the
five-dimensional Gödel universe.  Doing so we see that
\begin{equation*}
  P_0 + \cos\vartheta P_1 + \sin\vartheta T_1 - Y_0
\end{equation*}
does obey equation~\eqref{eq:geodetic} with $c=0$.  This means that
the plane-wave limits will again be regular.

Indeed, using equation~\eqref{eq:Fab}, we find that the only
nonzero components of $F$ are
\begin{equation*}
  F_{14} = - \half \sin\vartheta \qquad\text{and}\qquad F_{23} =
  -\half~.
\end{equation*}
Similarly, using equation~\eqref{eq:Aab} the matrix $A$ has nonzero
components
\begin{equation*}
  A_{11} = -1 + \tfrac34 \sin^2\vartheta~,\quad
  A_{22} = A_{33} = -\tfrac14~,\qquad\text{and}\qquad
  A_{44} = -\tfrac14 \sin^2\vartheta~.
\end{equation*}
Notice that since $[A,F]\neq 0$ this is not a symmetric plane wave.

\subsection{A lorentzian g.o.~space}
\label{sec:kaplan}

In this section we will discuss the geometry of a six-dimensional
lorentzian g.o.~space which is not naturally reductive.

\subsubsection{The geometry}

The lorentzian version of Kaplan's g.o.~space (see, for example,
\cite{KowNik}) is a six-dimensional 2-step nilpotent Lie group $M$
with a left invariant metric.  The Lie algebra $\fm$ is spanned by
$X_i$ for $i=1,\dots,6$ subject to the following nonzero Lie brackets:
\begin{equation}
  \label{eq:2nilalg}
  \begin{aligned}[m]
    [X_1,X_3] &= X_5\\
    [X_2,X_4] &= - X_5
  \end{aligned}
  \qquad
  \begin{aligned}[m]
    [X_1,X_4] &= X_6\\
    [X_2,X_3] &= X_6
  \end{aligned}
\end{equation}
and the left-invariant metric is induced from the inner product making
the $X_i$ a pseudo-orthonormal frame with $X_6$ timelike.  Notice that
this inner product is \emph{not} ad-invariant:
\begin{equation*}
  1 = \left<[X_1,X_3],X_5\right> \neq \left<X_1,[X_3,X_5]\right> = 0~,
\end{equation*}
whence the metric on the group is not bi-invariant.

Let us introduce a dual basis $\theta^i$ for $\fm^*$ which we extend
to left-invariant one-forms on the group $M$.  They obey the
Maurer--Cartan structure equation
\begin{equation*}
  d\theta^i(X,Y) = -\theta^i([X,Y])~,
\end{equation*}
whence $d\theta^i=0$ for $i=1,2,3,4$ and
\begin{equation*}
  d\theta^5 = - \theta^{13} + \theta^{24}
  \qquad\text{and}\qquad
  d\theta^6 = - \theta^{14} - \theta^{23}~,
\end{equation*}
where we have used the notation $\theta^{ij} = \theta^i \wedge \theta^j$.  To
integrate these equations, we introduce coordinate functions $x_i$
such that $\theta^i = dx_i$ for $i=1,2,3,4$ and
\begin{equation*}
  \theta^5 = dx_5 + x_3 dx_1 - x_4 dx_2
  \qquad\text{and}\qquad
  \theta^6 = dx_6 + x_4 dx_1 + x_3 dx_2,
\end{equation*}
relative to which the metric is given by
\begin{equation}
  \label{eq:Kaplan}
  \sum_{i=1}^4 dx_i^2 + (dx_5 + x_3 dx_1 - x_4 dx_2)^2 -
  (dx_6 + x_4 dx_1 + x_3 dx_2)^2~,
\end{equation}
which exhibits $M$ as an $\RR^2$-bundle over flat $\RR^4$, or
as a real line bundle over a five-dimensional Gödel metric
\begin{equation}
  \label{eq:Goedel}
  \sum_{i=1}^4 (dx^i)^2 - (dx_6 + x_4 dx_1 + x_3 dx_2)^2~,
\end{equation}
in different coordinates to the one in \eqref{eq:metric}.  In this
sense it is to be compared with the maximally supersymmetric plane
wave in six-dimensional $(1,0)$ supergravity \cite{Meessen} which is
also a line bundle over the Gödel metric (see, for example,
\cite{CFOSchiral}).  Parenthetically, this prompts the natural
question whether this six-dimensional g.o.~geometry can support flux
making it a homogeneous $(1,0)$ supergravity background.  The answer
to this question is negative.

\subsubsection{Isometries}

The Lie algebra of isometries of $M$ is a semidirect product $\fg =
\fh \ltimes \fm$, where $\fh$ consists of those (outer) derivations of
$\fm$ which are skew-symmetric relative to the inner product
$\left<-,-\right>$.  (In the riemannian case this follows from a
theorem of Gordon's \cite{Gordon}.)  Let $\delta$ be an outer
derivation.  Then $\delta$ preserves the centre $\fz$, which is the
span of $X_5,X_6$.  Since $\delta$ is skew-symmetric, it also
preserves the orthogonal complement $\fa = \fz^\perp$ of the centre,
spanned by $X_i$, $i=1,2,3,4$.  The Lie bracket in $\fm = \fa \oplus
\fz$ defines a map
\begin{equation*}
  \Lambda^2_+ \fa \to \fz
\end{equation*}
which is equivariant under the action of $\delta$.  It is not hard to
show that $\delta$ must in fact act trivially in both spaces, which
means that $\fh = \fso(\fa)_- \subset \fso(\fa)$ are anti-selfdual
rotations in $\fa$, whence $\fh \cong \fsu(2)$.  Let $Y_a$, $a=1,2,3$,
denote a basis for $\fh$.  Then the nonzero Lie brackets of $\fg$ are
given by \eqref{eq:2nilalg} together with
\begin{equation}
  \label{eq:KaplanIso}
  \begin{aligned}[m]
    [Y_1,X_1] &= X_3\\
    [Y_1,X_2] &= X_4\\
    [Y_1,X_3] &= -X_1\\
    [Y_1,X_4] &= -X_2\\[5pt]
    [Y_1,Y_2] &= -2Y_3
  \end{aligned}
  \qquad
  \begin{aligned}[m]
    [Y_2,X_1] &= X_4\\
    [Y_2,X_2] &= -X_3\\
    [Y_2,X_3] &= X_2\\
    [Y_2,X_4] &= -X_1\\[5pt]
    [Y_2,Y_3] &= -2Y_1
  \end{aligned}
  \qquad
  \begin{aligned}[m]
    [Y_3,X_1] &= X_2\\
    [Y_3,X_2] &= -X_1\\
    [Y_3,X_3] &= -X_4\\
    [Y_3,X_4] &= X_3\\[5pt]
    [Y_3,Y_1] &= -2Y_2
  \end{aligned}
\end{equation}

It is possible to write down the Killing vectors explicitly, in the
coordinates above.  First of all we notice that these coordinates are
such that the Maurer--Cartan one-form $\btheta = \sum_{i=1}^6 \theta^i
\otimes X_i$ is given by $\btheta = g(x)^{-1} dg(x)$, where
\begin{equation*}
  g(x) = \exp(x^1 X_1 + x^2 X_2) \exp(x^3 X_3 + x^4 X_4) \exp(x^5 X_5
  + x^6 X_6)~.
\end{equation*}
The Killing vectors in $\fm$ are the right-invariant vector fields on
$M$, since they generate left-translations.  Equivalently they are
dual to the right-invariant one-form
\begin{multline*}
  dg(x) g(x)^{-1} = \sum_{i=1}^4 dx^i \otimes X_i
  + (dx_5 + x_1 dx_3 - x_2 dx_4) \otimes X_5\\
  + (dx_6 + x_1 dx_4 + x_2 dx_3) \otimes X_6~;
\end{multline*}
that is,
\begin{equation*}
  \xi_{X_3} = \d_3 - x_1 \d_5 - x_2 \d_6
  \qquad\text{and}\qquad
  \xi_{X_4} = \d_4 + x_2 \d_5 - x_1 \d_6
\end{equation*}
and $\xi_{X_i} = \d_i$ for $i=1,2,5,6$, where $\d_i = \frac{\d}{\d
  x^i}$.  Notice that, as expected,
\begin{equation*}
  [\xi_{X_i}, \xi_{X_j}] = - \xi_{[X_i,X_j]}~.
\end{equation*}
The Killing vectors in $\fh$ are found by differentiating
\begin{equation*}
  \frac{d}{dt}\biggr|_0  \Ad(\exp(t Y)) g(x)~,
\end{equation*}
for any $Y \in \fh$.  Doing so we obtain
\begin{align*}
  \xi_{Y_1} &= -x_3 \d_1 - x_4 \d_2 + x_1 \d_3 + x_2 \d_4 - \half(
  x_1^2 - x_2^2 - x_3^2 + x_4^2) \d_5 - (x_1 x_2 - x_3 x_4) \d_6\\
  \xi_{Y_2} &= -x_4 \d_1 + x_3 \d_2 - x_2 \d_3 + x_1 \d_4 - (x_1 x_2 +
  x_3 x_4) \d_5 + \half(x_1^2 - x_2^2 + x_3^2 - x_4^2) \d_6\\
  \xi_{Y_3} &= -x_2 \d_1 + x_1 \d_2 + x_4 \d_3 - x_3 \d_4~.
\end{align*}
Again it can be checked that as expected, the map $X \mapsto \xi_X$,
for $X\in\fg$, is a Lie algebra anti-homomorphism: $[\xi_X,\xi_Y] = -
\xi_{[X,Y]}$.

\subsubsection{Null geodesics}

It is easy to describe the null geodesics on $M$ up to isometry.  By
homogeneity we can have them pass by any point, in particular the
identity of $M$, thought of as a Lie group.  The tangent vector is
then a null vector $\boldsymbol{v} = \sum_{i=1}^6 v^i X_i \in \fm$,
with $\sum_{i=1}^5 v_i^2 = v_6^2$.  We can choose without loss of
generality $v_6 = \pm 1$, depending on whether it is future- or
past-pointing, respectively.  The isotropy group $\SU(2) \subset
\SO(4)$ leaves $X_5$ invariant and acts transitively on the spheres in
the orthogonal four-dimensional space. This means that up to isometry,
there is a (quarter-)circle family of past- and future-pointing null
geodesics, with tangent vectors
\begin{equation}
  \label{eq:nullgeodetic}
  \boldsymbol{v} = \sin\vartheta X_1 + \cos\vartheta X_5 \pm X_6~,
\end{equation}
for $\vartheta \in [0,\frac\pi2]$.

\subsubsection{Geodesic orbits}

The geodesic orbit nature of this homogeneous space is easy to see.
Remember that this requires finding, for every $0 \neq X\in\fm$ a
$\phi(X) \in \fh$ such that $X + \phi(X)$ is geodetic; that is,
such that equation~\eqref{eq:graph} is satisfied for all
$Z\in\fm$.  One finds that if $X = \sum_{i=1}^6 v_i X_i$, then
letting $\phi(X) = \sum_{a=1}^3 \phi_a Y_a$, where
\begin{align*}
  \phi_1 &= (v_1^2 - v_2^2 + v_3^2 - v_4^2)
  \frac{v_5}{|\boldsymbol{\bv}_\perp|^2} - 2 (v_1 v_2 + v_3
  v_4)\frac{v_6}{|\boldsymbol{\bv}_\perp|^2}\\
  \phi_2 &= -2(v_1 v_2 - v_3 v_4) \frac{v_5}{|\boldsymbol{\bv}_\perp|^2} -
  (v_1^2 - v_2^2 - v_3^2 + v_4^2)\frac{v_6}{|\boldsymbol{\bv}_\perp|^2}\\
  \phi_3 &= 2 (v_1 v_4 + v_2 v_3) \frac{v_5}{|\boldsymbol{\bv}_\perp|^2} + 2 (
  v_1 v_3 -  v_2 v_4) \frac{v_6}{|\boldsymbol{\bv}_\perp|^2}~,
\end{align*}
where $|\boldsymbol{\bv}_\perp|^2 = \sum_{i=1}^4 v_i^2$, yields a
geodetic vector with $c=0$ in equation \eqref{eq:graph}.  Notice that
for the null geodesic with tangent vector $\bv$ given by
\eqref{eq:nullgeodetic}, we find
\begin{equation*}
  \phi_1 = v_5 = \cos\vartheta \qquad \phi_2 = -v_6 = \mp 1\qquad
  \phi_3 = 0~,
\end{equation*}
which is the restriction of a linear function.  In other words,
whereas $\phi: \fm\backslash \{0\} \to \fh$ is nonlinear (showing that
$M$ is not naturally reductive), it is in some sense like a naturally
reductive space when restricted to (certain) null geodesics.

\subsubsection{Plane-wave limits}

Let us consider the geodetic vector
\begin{equation*}
  \sin\vartheta X_1 + \cos\vartheta X_5 \pm X_6 + \cos\vartheta Y_1
  \mp Y_2~.
\end{equation*}
Using equation~\eqref{eq:Fab}, we find that
\begin{equation*}
  F =
  \begin{pmatrix}
    0 & 0 & -\half & \pm \half \cos\vartheta\\
    0 & 0 & \mp\frac32 & -\frac32 \cos\vartheta\\
    \half & \pm \frac32 & 0 & 0\\
    \mp\half\cos\vartheta &  \frac32 \cos\vartheta & 0 & 0
  \end{pmatrix}~,
\end{equation*}
and, using \eqref{eq:Aab}, that
\begin{equation*}
  A =
  \begin{pmatrix}
    \frac18 (-3 - \cos2\vartheta) & \mp \frac34 \sin^2\vartheta & 0 &  0 \\
    \mp\frac34 \sin^2\vartheta & \frac18 (-3-\cos2\vartheta) & 0 & 0 \\
    0 & 0 & -\half \cos2\vartheta & 0 \\
    0 & 0 & 0 & \frac14 (-3 + 2\cos2\vartheta)
  \end{pmatrix}~.
\end{equation*}
It is easy to see that $[A,F] = 0$ if and only if $\vartheta=0$, in
which case the resulting spacetime is a conformally flat symmetric
plane wave.

\subsubsection{Homogeneous structures}

We start with the reductive split $\fg = \fh \oplus \fm$ with $\fg$
the full isometry algebra.  The resulting homogeneous structure can be
calculated using equation \eqref{eq:S@oexplicit}.  Doing so we find a
homogeneous structure of type $\eT_2 \oplus \eT_3$, with components
$S_{ijk} = S(X_i,X_j,X_k)$ given by
\begin{gather*}
  S_{135} = S_{326} = S_{416} = S_{425} = S_{524} = S_{614} = S_{623}
  = \half\\
  S_{146} = S_{236} = S_{245} = S_{315} = S_{513} = - \half~.
\end{gather*}
As explained at the end of Section~\ref{sec:structures}, we can search
for other homogeneous structures by restricting to subalgebras $\fg'
\subseteq \fg$ and looking for reductive splits $\fg' = \fh' \oplus
\fm'$, where $\fh'= \fg' \cap \fh$ and $\fm'$ is the graph of an
$\fh'$-equivariant linear map $\fm \to \fh'$.

First of all we notice that there are no nontrivial $\fh$-equivariant
linear maps $\fm \to \fh$, since decomposing $\fh$ and $\fm$ into
irreducible $\fh$-modules, we see that they have no isotypical
submodules in common: $\fh$ is simple, whence irreducible and
three-dimensional, whereas $\fm$ breaks up into two one-dimensional
trivial submodules and an irreducible four-dimensional submodule.
Therefore to obtain other homogeneous structures, we must consider
proper subalgebras $\fg' \subsetneq \fg$.  It is only necessary to
consider subalgebras up to conjugation, whence there is only one
possibility: any one-dimensional subalgebra $\fh' \subset \fh$, e.g.,
the one spanned by $Y_1$, say.  Any other choice is related by
conjugation and will give rise to isomorphic homogeneous structures.

Decomposing $\fm$ and $\fh'$ into irreducible representations of
$\fh'$ we find
\begin{equation*}
  \fm = \RR_0  \oplus \RR_0 \oplus \RR^2_1 \oplus \RR^2_1
  \qquad\text{and}\qquad \fh' = \RR_0~,
\end{equation*}
where the subscripts indicate the highest weight of the
representation.  The trivial representations in $\fm$ are spanned by
$X_5$ and $X_6$, respectively, whereas the two-dimensional
representations are spanned by $X_1,X_3$ and $X_2,X_4$, respectively.
We therefore have a two-parameter family of $\fh'$-equivariant linear
maps $\varphi: \fm \to \fh'$, given by
\begin{equation*}
  \varphi (v^i X_i) = (\alpha v^5 + \beta v^6) Y_1~.
\end{equation*}
The graph of $\varphi$ is then the subspace $\fm'\subset \fg' =
\fh'\oplus \fm$ spanned by
\begin{equation*}
  X_1~, \quad X_2~, \quad X_3~, \quad X_4~, \quad X_5 + \alpha Y_1~,
  \quad\text{and}\quad  X_6 + \beta Y_1~.
\end{equation*}
This means that the $[-,-]_{\fm'}$ brackets change; for example,
\begin{equation*}
  \begin{aligned}[t]
    [X_5 + \alpha Y_1, X_1]_{\fm'} &= \alpha X_3\\
    [X_5 + \alpha Y_1, X_2]_{\fm'} &= \alpha X_4\\
    [X_5 + \alpha Y_1, X_3]_{\fm'} &= - \alpha X_1\\
    [X_5 + \alpha Y_1, X_4]_{\fm'} &= - \alpha X_2
  \end{aligned}
  \qquad\qquad
  \begin{aligned}[t]
    [X_6 + \beta Y_1, X_1]_{\fm'} &= \beta X_3\\
    [X_6 + \beta Y_1, X_2]_{\fm'} &= \beta X_4\\
    [X_6 + \beta Y_1, X_3]_{\fm'} &= - \beta X_1\\
    [X_6 + \beta Y_1, X_4]_{\fm'} &= - \beta X_2~.
  \end{aligned}
\end{equation*}
We can now compute the corresponding homogeneous structure using
formula \eqref{eq:S@oexplicit} and we obtain a two-parameter family of
$\eT_2 \oplus \eT_3$ structures:
\begin{gather*}
  S_{326} = S_{416} = S_{614} = S_{623} = \half\\
  S_{146} = S_{246} = -\half\\
  S_{316} = S_{426} = S_{613} = S_{624} = \half\beta\\
  S_{136} = S_{246} = -\half\beta\\
  S_{135} = \half (1 + \alpha)\\
  S_{245} = - \half (1 - \alpha)\\
  S_{315} = S_{513} = - \half (1 + \alpha)\\
  S_{425} = S_{524} = \half (1 - \alpha)~.
\end{gather*}
Naturally, when $\alpha = \beta = 0$ we recover the homogeneous
structure of the maximal reductive split.

It is instructive to compare this with the explicit solution of the
Ambrose--Singer equations \eqref{eq:AmbroseSinger}.  Solving these
equations gives a general solution labelled by six parameters
$z_1,\ldots,z_6$ in the intersection of three quadrics:
\begin{equation*}
  z_1 z_5 = z_2 z_4 \qquad
  z_1 z_6 = z_3 z_4 \qquad
  z_3 z_5 = z_2 z_6~.
\end{equation*}
These equations are equivalent to the matrix
\begin{equation*}
  \begin{pmatrix}
    z_1 & z_2 & z_3\\
    z_4 & z_5 & z_6
  \end{pmatrix}
\end{equation*}
having rank $<2$.  The general solution of such equations is given in
terms of two vectors $\bv = (v_1,v_2) \in \RR^2$ and $\bw =
(w_1,w_2,w_3) \in \RR^3$, by
\begin{equation*}
  \begin{pmatrix}
    z_1 & z_2 & z_3\\
    z_4 & z_5 & z_6
  \end{pmatrix} =
  \begin{pmatrix}
    v_1 w_1 & v_1 w_2 & v_1 w_3\\
    v_2 w_1 & v_2 w_2 & v_2 w_3
  \end{pmatrix}~.
\end{equation*}
It is not hard to show that two such homogeneous structures labelled
by $(\bv, \bw)$ and $(\bv', \bw')$ are isomorphic if and only if $\bw$
and $\bw'$ are related by an $\SO(3) = \Ad \SU(2)$ transformation.
Any $\bw \in \RR^3$ is $\SO(3)$-related to $(w_1,0,0)$, in which case
the solution has only two parameters $v_1w_1$ and $v_2w_1$,
corresponding to our $\alpha$ and $\beta$ above.

\subsection{Komrakov spacetimes}

In \cite{Komrakov} there is a classification of four-dimensional
pseudo-riemannian (locally) homogeneous spaces.  The Komrakov list is
a useful source of examples on which to test conjectures.  In this
section we will present two of them to illustrate the discussion in
the bulk of the paper.  The nomenclature follows \cite{Komrakov}.

\subsubsection{Komrakov 1.1$^2_{\lambda=0}$}
\label{sec:Komrakov112}

This case in Komrakov's classification has a parameter ($\lambda$)
which we are putting to zero in order for the resulting homogeneous
space to admit a lorentzian metric.  For $\lambda\neq 0$ the metric is
either riemannian or hyperbolic.  The isometry algebra is a
semidirect product $\fg = \fh \ltimes \fm$ with $\fh$
one-dimensional with basis $e_1$ and $\fm$ four-dimensional with basis
$u_1,\ldots,u_4$.  The nonzero Lie brackets are
\begin{equation*}
  \begin{aligned}[t]
    [u_4,u_1]&=-u_1\\
    [u_4,u_2]&=-2u_2\\
    [e_1,u_1]&=u_3
  \end{aligned}
  \qquad\qquad
  \begin{aligned}[t]
    [u_4,u_3]&=-u_3\\
    [u_1,u_3]&=-u_2\\
    [e_1,u_3]&=-u_1
  \end{aligned}~.
\end{equation*}
Up to homothety (and Lie algebra automorphisms), there is a
two-parameter family of $\fh$-invariant lorentzian metrics
$\left<u_1,u_1\right> = \left<u_3,u_3\right> = 1$,
$\left<u_2,u_2\right> = \alpha$ and $\left<u_4,u_4\right> = \beta$,
with $\alpha\beta < 0$.

The homogeneous structure corresponding to this split is given by
equation \eqref{eq:S@oexplicit} and has (nonzero) components $S_{ijk}
= S(u_i,u_j,u_k)$ given by
\begin{equation*}
  S_{123} = S_{213} = S_{312} = \half \alpha \qquad S_{224} = -2\alpha
  \qquad S_{114} = S_{334} = -1~,
\end{equation*}
which is of generic type $\eT_1 \oplus \eT_2 \oplus \eT_3$.

It is possible to deform this homogeneous structure by choosing a
different reductive split $\fg = \fh \oplus \fm'$ with $\fm'$ the
graph of an $\fh$-equivariant linear map $\varphi: \fm \to \fh$.  We
find that there is a 2-parameter family of such maps, and hence a
2-parameter family of such splits.  Indeed, let $\fm'$ denote the span
of the following vectors
\begin{equation*}
  u_1~, \quad u_2 + c_2 e_1~, \quad  u_3~, \quad\text{and}\quad
  u_4 + c_4 e_1~,
\end{equation*}
with resulting homogeneous structure
\begin{gather*}
  S_{123} = S_{312} = \half \alpha \qquad S_{213} = c_2 + \half \alpha
  \qquad S_{224} = -2\alpha\\
  S_{114} = S_{334} = -1 \qquad S_{413} = c_4~.
\end{gather*}
For generic values of $c_2,c_4$ this is again of type $\eT_1 \oplus
\eT_2 \oplus \eT_3$, but there is a point, $c_2 = \half\alpha$ and
$c_4=0$, for which the $\eT_3$ component is absent.

Up to the action of the isotropy, a null vector (at the identity coset)
can be written as
\begin{equation*}
  v^1 u_1 + v^2 u_2 + v^4 u_4~,
\end{equation*}
where $(v^1)^2 + \alpha (v^2)^2 + \beta (v^4)^2 = 0$.  We must
distinguish between two cases: $\alpha<0, \beta>0$ and
$\alpha>0,\beta<0$.  In either case, the timelike component can be set
to $1$ (for future-pointing null rays) without loss of generality.

\begin{itemize}
\item $\alpha<0,\beta>0$.\\
  In this case, the null vector is $u_2 + p u_4 + q u_1$, with $q =
  \sqrt{-\alpha - \beta p^2}$.  We find that the geodetic equation
  \eqref{eq:geodetic} has a unique solution, with geodetic vector
  \begin{equation*}
    u_2 + p u_4 \qquad \text{with $p^2=-\alpha/\beta$ and $c=-2p$}.
  \end{equation*}
  The plane-wave limit along this homogeneous geodesic will give rise
  to a singular homogeneous plane wave.

\item $\alpha>0,\beta<0$.\\
  In this case, the null vector is $u_4 + p u_2 + q u_1$, with $q =
  \sqrt{-\beta-\alpha p^2}$.  Here we find two homogeneous geodesics:
  \begin{align*}
    & u_4 + p u_2 \qquad \text{with $p^2=-\beta/\alpha$ and $c=-2$},\\
    & u_4 + q u_1 \qquad \text{with $q^2=-\beta$ and $c=-1$}.
  \end{align*}
  Again the corresponding plane-wave limits will give rise to singular
  homogeneous plane waves.
\end{itemize}

For ease of exposition we will take $|\alpha| = |\beta|=1$ from now
on.  We consider three cases:
\begin{itemize}
\item $v = u_2 \pm u_4$, $c=\mp 2$, $\alpha = -1$, $\beta=1$;\\
  In this case, the skew-symmetric matrix $F$ has components $F_{12} =
  \half$, whereas the symmetric matrix $A$ is given by
  \begin{equation*}
    A =
    \begin{pmatrix}
      \frac34 & \pm 1 \\ \pm 1 & \frac34
    \end{pmatrix}~.
  \end{equation*}
  It is clear that $[F,A] \neq 0$.  Indeed,
  \begin{equation*}
    e^{z F} A e^{-z F} =
    \begin{pmatrix}
      \frac34 \pm \sin z & \pm \cos z\\
      \pm \cos z & \frac34 \mp \sin z
    \end{pmatrix}~.
  \end{equation*}

\item $v = u_4 \pm u_2$, $c=-2$, $\alpha =1$, $\beta=-1$\\
  In this case, the skew-symmetric matrix $F$ has components $F_{12} =
  \mp\half$, whereas the symmetric matrix $A$ is given by
  \begin{equation*}
    A =
    \begin{pmatrix}
      \frac34 & \mp 1 \\ \mp 1 & \frac34
    \end{pmatrix}~.
  \end{equation*}
  Again, it is clear that $[F,A] \neq 0$.

\item $v = u_4 \pm u_1$, $c=1$, $\alpha =1$, $\beta=-1$\\
  Finally, in this case, the skew-symmetric matrix $F$ has components
  $F_{12} = \mp\half$, whereas the symmetric matrix $A$ is given by
  \begin{equation*}
    A =
    \begin{pmatrix}
      \frac{15}4 & \pm 1 \\ \pm 1 & \frac74
    \end{pmatrix}~.
  \end{equation*}
  Again $[A,F]\neq 0$ and indeed
  \begin{equation*}
    e^{z F} A e^{-z F} =
    \begin{pmatrix}
      \frac{11}4 + \cos z - \sin z & \pm (\cos z + \sin z)\\
      \pm (\cos z + \sin z) & \frac{11}4 - \cos z + \sin z
    \end{pmatrix}~.
  \end{equation*}
\end{itemize}

\subsubsection{Komrakov 1.4.6}
\label{sec:komrakov146}

The isometry algebra is the semidirect product $\fg = \fh \ltimes \fm$
of a one-dimensional Lie algebra $\fh$ spanned by $e_1$ and a
four-dimensional Lie algebra spanned by $u_1,\ldots,u_4$.
The nonzero Lie brackets are
\begin{equation*}
  \begin{aligned}[t]
    [u_1,u_4]&=u_1\\
    [u_2,u_4]&=u_2\\
    [u_3,u_4]&=u_1 + u_3
  \end{aligned}
  \qquad\qquad
  \begin{aligned}[t]
    [e_1,u_2]&=u_1\\
    [e_1,u_3]&=u_2~.
  \end{aligned}
\end{equation*}
Up to homothety (and Lie algebra automorphism) there is a unique
$\fh$-invariant lorentzian inner product on $\fm$:
$\left<u_1,u_3\right> = -1$ and $\left<u_2,u_2\right> =
\left<u_4,u_4\right> = 1$.

There is a two-parameter family of $\fh$-equivariant linear map $\fm
\to \fh$.  The graph $\fm'$ of a map in this family (labelled by
$\alpha$ and $\beta$) is the subspace of $\fg$ spanned by
\begin{equation*}
  u_1 + \alpha e_1~, \quad u_2 ~, \quad  u_3~, \quad\text{and}\quad
  u_4 + \beta e_1~.
\end{equation*}

The subspace $\fm'$ is no longer a Lie subalgebra, but projecting the
brackets to $\fm'$ we obtain
\begin{equation*}
  \begin{aligned}[t]
    [u_1+\alpha e_1,u_4+\beta e_1]_{\fm'}&=u_1+\alpha e_1\\
    [u_1+\alpha e_1,u_3]_{\fm'}&= \alpha u_2\\
    [u_1+\alpha e_1,u_2]_{\fm'}&= \alpha (u_1+\alpha e_1)\\
  \end{aligned}
  \qquad\qquad
  \begin{aligned}[t]
    [u_2, u_4 + \beta e_1]_{\fm'}&=u_2 - \beta (u_1 + \alpha e_1)\\
    [u_3, u_4 + \beta e_1]_{\fm'}&=u_1 + u_3 - \beta u_2~.
  \end{aligned}
\end{equation*}

The resulting homogeneous structure has components $S_{ijk} =
S(u_i,u_j,u_k)$ given by
\begin{gather*}
  S_{134} = S_{314} = S_{334} = 1 \qquad S_{123}= -\alpha \qquad
  S_{423} = -\beta \qquad S_{224} = -1~,
\end{gather*}
which is generically of type $\eT_1 \oplus \eT_2 \oplus \eT_3$, but of
type $\eT_1 \oplus \eT_2$ when $\alpha = \beta = 0$.

To determine the plane-wave limits along homogeneous geodesics, we
first determine the null directions up to the action of isometries.
Let $v = \sum_i v^i u_i \in \fm$ be a null vector.  Then
\begin{equation*}
  2 v^1 v^3 = (v^2)^2 + (v^4)^2~.
\end{equation*}
The action of the isotropy is obtained by exponentiating the adjoint
action of $e_1\in\fh$:
\begin{equation*}
  \begin{pmatrix}
    v^1 \\ v^2 \\ v^3 \\ v^4
  \end{pmatrix}
  =
  \begin{pmatrix}
    1 & t & \half t^2 & 0 \\
    0 & 1 & t & 0\\
    0 & 0 & 0 & 0\\
    0 & 0 & 0 & 0
  \end{pmatrix}
  \begin{pmatrix}
    v^1 \\ v^2 \\ v^3 \\ v^4
  \end{pmatrix}
  =
  \begin{pmatrix}
    v^1 + t v^2 + \half t^2 v^3 \\
    v^2 + t v^3\\
    v^3\\
    v^4
  \end{pmatrix}~.
\end{equation*}
We must distinguish between two cases:
\begin{itemize}
\item If $v^3 = 0$ then so are $v^2$ and $v^4$, whereas $v^1\neq 0$.
  Therefore the null vector can be chosen to be $u_1$.
\item If $v^3 \neq 0$ then we can use the isotropy action to put
  $v^2=0$ and rescale the null vector to make $v^3 = 1$, so that the
  null vector is then $u_3 + \alpha u_4 + \half \alpha^2 u_1$ for some
  $\alpha \in \RR$.
\end{itemize}
A simple calculation shows that in the first case, $u_1$ satisfies
equation~\eqref{eq:geodetic} with $c=0$ when the plane-wave limit
along that geodesic will be regular.  In contrast, for the second
case, there is no value of $\alpha$ for which the corresponding
geodesic is homogeneous.

It is now a simple exercise to use equations~\eqref{eq:Fab} and
\eqref{eq:Aab} to show that the plane-wave limit along $u_1$ is
actually flat.

\section*{Acknowledgments}

We are grateful to Harry Braden for making us aware of
\cite{TricerriVanhecke} and to José Antonio Oubiña for some early
useful correspondence.

JMF would like to thank Paul Gauduchon and le Centre de Mathématiques
Laurent Schwartz de l'École Polytechnique, as well as il Centro
Culturale Merzagora--Petrini for their hospitality during the final
stages of this work.

PM would like to thank Matthias Blau, Bert Janssen and Tomás Ortín for
fruitful discussions and Josefina Millán Jiménez for her (library)
support.  PM would also like to thank el Instituto de Física Teórica
(Madrid) and het Instituut voor Theoretische Fysica (Leuven) for their
support and hospitality.

The research of JMF was partially funded in its initial stages by the
EPSRC grant GR/R62694/01 and that of SP by an EPSRC Postgraduate
studentship.

Finally, JMF would like to dedicate this paper to the memory of
Stanley Hobert, who used to teach Physics at the Pasadena City
College, and of whose untimely and tragic death he has recently become
aware.  Stanley Hobert was one of the many unsung heroes whose
contagious enthusiasm for Physics has inspired many a scientific
career.  JMF is personally in his debt and would like to hereby
acknowledge his gratitude and at the same time honour his memory.

\bibliographystyle{utphys}
\bibliography{AdS,AdS3,ESYM,Sugra,Geometry,CaliGeo}

\end{document}